\documentclass{article}
\pdfoutput=1
\usepackage{jcappub}
\usepackage{amsmath}
\usepackage{amssymb}
\usepackage{hyperref}
\usepackage{multirow}
\renewcommand{\vec}[1]{\boldsymbol{#1}}

\title{Impact of freeze-in on dark matter isocurvature}

\author[a]{N. Bellomo,}
\emailAdd{nicola.bellomo@austin.utexas.edu}

\author[b]{Kim V. Berghaus,}
\emailAdd{kim.berghaus@stonybrook.edu}

\author[a]{Kimberly K. Boddy}
\emailAdd{kboddy@physics.utexas.edu}

\affiliation[a]{Texas Center for Cosmology and Astroparticle Physics, Weinberg Institute, Department of Physics, The University of Texas at Austin, Austin, TX 78712, USA.}
\affiliation[b]{C. N. Yang Institute for Theoretical Physics, Stony Brook University, Stony Brook, NY 11794, USA.}

\abstract{
Dark matter freeze-in is a compelling cosmological production mechanism in which all or some of the observed abundance of dark matter is generated through feeble interactions it has with the Standard Model.
In this work we present the first analysis of freeze-in dark matter fluctuations and consider two benchmark models: freeze-in through the direct decay of a heavy vector boson and freeze-in through pair annihilation of Standard Model particles in the thermal bath.
We provide a theoretical framework for determining the impact of freeze-in on curvature and dark matter isocurvature perturbations.
We determine freeze-in dark matter fluid properties from first principles, tracking its evolution from its relativistic production to its final cold state, and calculate the evolution of the dark matter isocurvature perturbation.
We find that in the absence of initial isocurvature, the freeze-in production of dark matter does not source isocurvature.
However, for an initial isocurvature perturbation seeded by inflation, the nonthermal freeze-in process may allow for a fraction of the isocurvature to persist, in contrast to the exponential suppression it receives in the case of thermal dark matter.
In either case, the evolution of the curvature mode is unaffected by the freeze-in process.
We show sensitivity projections of future cosmic microwave background experiments to the amplitude of uncorrelated, totally anticorrelated, and totally correlated dark matter isocurvature perturbations.
From these projections, we infer the sensitivity to the abundance of freeze-in dark matter that sustains some fraction of the primordial isocurvature.
}

\begin{document}

\maketitle

\section{Introduction}
\label{sec:intro}

Cosmological and astrophysical observations provide incontrovertible evidence for the existence of dark matter (DM)~\cite{1970ApJ...159..379R, Komatsu:2008hk, aghanim:planck2018cosmoparameters}.
A cold and collisionless DM component of the Universe describes the large-scale structure of the Universe remarkably well, as evidenced by the anisotropy of the cosmic microwave background (CMB)~\cite{aghanim:planck2018cosmoparameters} and galaxy clustering~\cite{demattia:bossclustering, chapman:ebossclustering}, for example.

One of the prominent theoretical descriptions for the nature of DM is the weakly-interacting massive particle (WIMP), for which DM has interactions with Standard Model (SM) particles beyond gravity~\cite{arcadi:wimpreview}.
These interactions are weak enough for WIMPs to be considered collisionless for the purposes of structure formation and yet substantial enough for WIMPs to be in equilibrium with the bath of thermalized SM particles at very early times in cosmic history.
As the Universe expands and cools, WIMPs eventually undergo thermal freeze-out, in which the comoving number density approaches a fixed value after chemical decoupling from the SM bath.

A compelling alternative scenario for the cosmological production of DM is freeze-in~\cite{Hall:2009bx}.
For instance, the annihilation of SM particles can generate an abundance of feebly-interacting massive particles constituting DM, but the interaction strength is so weak that DM particles never attain thermal equilibrium with the SM bath~\cite{Hall:2009bx}. 
Alternatively, the feeble decay of a heavy parent particle thermalized with the SM bath could generate the observed DM abundance without the DM attaining thermal equilibrium.
Beyond these examples there exists a variety of well-motivated models that can give rise to freeze-in dynamics in the early Universe that are compatible with the observed properties of DM~\cite{Hall:2009bx, Chu:2011be, Blennow:2013jba, Bernal:2017kxu, DEramo:2017ecx, Belanger:2018ccd, Heeba:2018wtf, Belanger:2020npe, dvorkin:freezeinI, dvorkin:freezeinII, Elor:2021swj, Fernandez:2021iti}.

On the experimental side, freeze-in DM sets an important cosmological benchmark for direct detection experiments searching for evidence of DM interactions via electronic 
recoils~\cite{Essig:2011nj, Derenzo:2016fse,  Battaglieri:2017aum, PhysRevLett.121.051301, PhysRevLett.121.111303, PhysRevLett.121.061803, PhysRevLett.122.161801, PhysRevLett.123.181802, PhysRevLett.123.251801, PhysRevD.102.091101, PhysRevLett.125.171802, PhysRevLett.125.141301, PhysRevLett.126.211803, essig:snowmass2021}.
Typical WIMP-motivated nuclear-recoil experiments lose sensitivity for DM masses below $\mathcal{O}(\mathrm{GeV})$, while electronic-recoil experiments can probe DM masses down to $\mathcal{O}(\mathrm{MeV})$.
For low masses, thermal WIMPs generate too much relativistic energy density during big bang nucleosynthesis (BBN), altering predictions of the primordial element abundances~\cite{kolb:wimpeffectonbbn, serpico:wimpeffectonbbn, boehm:wimpeffectonbbn}.
Freeze-in provides a mechanism to produce such light DM candidates without running afoul of BBN constraints~\cite{baumholzer:freezeinNeff, ballesteros:freezeinNeff, li:freezeinNeff, decant:freezeinNeff}.

The nonthermal nature of freeze-in production has been studied in the context of their nonthermal phase space distribution (PSD) function and its impact on cosmological observables \cite{Bae:2017dpt, dvorkin:freezeinI, dvorkin:freezeinII, Du:2021jcj, Huang:2023jxb}.
These previous works considered the impact of out-of-equilibrium production on only the background evolution of the Universe.

In this work we determine the properties of DM fluctuations during freeze-in and how they influence the evolution of a primordial DM isocurvature perturbation.
While single field inflation fits CMB observations remarkably well \cite{chowdury:planckjudgementday} and only allows for adiabatic perturbations, multi-field inflation is a theoretically well-motivated alternative that can give rise to an initial isocurvature~\cite{wands:multifieldinflationbook}. 
However, since isocurvature can evolve in a multi-fluid system that exhibits energy exchange, the primordial value is not necessarily the one that is constrained by cosmological observables.
This effect has been studied explicitly in the context of several scalar fields that exchange energy with one another (e.g.\ \cite{Malik:2004tf}). 
DM freeze-in offers an alternative scenario of a post-inflationary process that exhibits energy exchange between sectors, which do not reach thermal equilibrium, and thus does not automatically wash out information about the state of an initial abundance of DM just after inflation.  
In our work we study for the first time the impact of DM freeze-in on the evolution of the adiabatic and isocurvature modes.

We focus on two DM freeze-in benchmark models, direct decay of a heavy parent particle into two DM particles and 2-to-2 annihilation of SM particles into millicharged DM, described in section~\ref{subsec:models}. 
We define a theoretical framework that encompasses the evolution of both the DM background phase space distribution (PSD) and its fluctuations from first principles in sections~\ref{subsec:background_cosmology} and~\ref{subsec:perturbative_cosmology}, following the evolution of both the adiabatic and isocurvature modes through the entire freeze-in process.
We use the PSD to compute both DM background and perturbed fluid properties, which we then use to investigate the evolution of the DM isocurvature perturbation mode during DM freeze-in in section~\ref{sec:kodama-sasaki_formalism}.
We find that, in presence of an initial fraction of DM characterized by an initial isocurvature perturbation, the process of DM freeze-in leads to a suppression of the initial isocurvature but does not exponentially wash it away as in the case of thermal DM~\cite{weinberg:adiabaticinitialconditions}.
On the other hand, in absence of a primordial isocurvature perturbation, no isocurvature is produced by DM formation.
Lastly, in section~\ref{sec:detectability_forecast} we forecast the sensitivity of next-generation CMB experiments to DM isocurvature perturbations and determine the corresponding sensitivity to the maximum initial DM fraction with isocurvature perturbations that can be generated during inflation, assuming the remaining DM abundance is generated via freeze-in.
Although we focus on freeze-in models in this paper, our roadmap is more general and can be applied to other DM formation mechanisms.
 

\section{Freeze-in cosmology}
\label{sec:freeze-in}

The evolution of the DM sector is described by the Boltzmann equation for the DM PSD~$f_\chi$:
\begin{equation}
    \frac{\partial f_{\chi}}{\partial \tau} + \frac{d x^i}{d \tau} \frac{\partial f_{\chi}}{\partial x^i} + \frac{d(ap)}{d\tau} \frac{\partial f_{\chi} }{\partial (a p)} = \frac{a(\tau)}{E} C[f_{i}] \,  ,  
\label{eq:boltzmann_equation}
\end{equation}
keeping terms up to linear order in perturbation theory, where~$\tau$ is the conformal time, $a(\tau)$ is the scale factor, $x^j$ are comoving spatial coordinates, and $m_\chi$, $p$, and~$E = \sqrt{p^2 + m_\chi^2}$ are the DM mass, momentum, and energy, respectively.
The collision term~$C[f_i]$ sources the production of DM from particle species~$i$, with PSD~$f_i$, in the thermal bath.
In this work we choose to evolve the PSD directly, which has the advantage of not requiring the introduction of additional equations to close the system, in contrast to, for instance, solving Friedmann equations where an equation of state for the fluids has to be assumed.
Instead, any non-trivial dynamical evolution of macroscopic DM properties can be exactly derived from a PSD approach, as we show in sections~\ref{subsec:background_cosmology} and~\ref{subsec:perturbative_cosmology}.

Typically, eq.~\eqref{eq:boltzmann_equation} is solved only at the background level for freeze-in dynamics; however, for the purpose of determining the evolution of isocurvature perturbations, we go one step further and study the dynamics of perturbations.
In other words, first we expand both the DM PSD and the collision term as
\begin{align}
    f_{\chi}(\vec{x},\vec{p},\tau) &= \bar{f}_{\chi}(\vec{p},\tau) + \delta f_\chi(\vec{x},\vec{p},\tau) \,, \\
    C(\vec{x},\vec{p},\tau) &= \bar{C}(\vec{p},\tau) + \delta C (\vec{x},\vec{p},\tau) \,,
\end{align}
where~$\bar{f}_{\chi}$ and~$\bar{C}=C[\bar{f_i}]$ are the homogeneous and isotropic background components with no spatial dependence, and $\delta f_\chi \ll \bar{f}_\chi$ and~$\delta C \ll \bar{C}$ are linear perturbations, which contains spatial fluctuations.
We then proceed by solving separately eq.~\eqref{eq:boltzmann_equation} at the background and perturbative level to determine both~$\bar{f}_\chi$ and~$\delta f_\chi$.

Throughout this study we approximate the early Universe to have two components: the DM fluid and the SM bath, consisting of all relativistic SM particles in thermal equilibrium at temperature $T$.
We denote the SM bath sector with the subscript ``$\mathrm{r}$'' for radiation.
This description of SM particles all in equilibrium at a single temperature breaks down when neutrinos decouple around $T \sim 1\, \text{MeV}$; thus, we restrict our study to DM freeze-in scenarios that conclude before neutrino decoupling.
The energy density of the SM bath is 
\begin{equation}
    \rho_\mathrm{r} = \frac{\pi^2}{30} g_*(T) T^4 \, ,
    \label{eq:energy-density-bath}
\end{equation}
where $g_*(T)$ is the effective number of relativistic degrees of freedom at temperature $T$.
Any particle species in kinetic equilibrium has a PSD given by a Bose-Einstein or Fermi-Dirac distribution
\begin{equation}
    f_i = \frac{1}{e^{E_i/T}\pm 1} \, ,
\end{equation}
depending on the spin statistics of the particle ($-$ for bosons, $+$ for fermions), where $E_i$ is the particle's energy.


\subsection{Dark matter production models}
\label{subsec:models}

The collision term appearing on the right-hand side of eq.~\eqref{eq:boltzmann_equation} dictates how the DM sector is created.
Here we analyze two representative DM freeze-in channels: DM production through the direct decay of a heavy parent particle thermalized with the SM bath and through the annihilation of SM particles.
For both cases, we assume DM is a spin-$1/2$ fermion.
The collision terms reported in section \ref{subsec:models} are valid at all order in perturbation theory, and we further specify them at the background and perturbative level in sections~\ref{subsec:background_cosmology} and~\ref{subsec:perturbative_cosmology}, respectively.


\subsubsection{Direct decay}
\label{subsubsec:direct_decay}

For the case of direct decay, we consider a vector particle~$Z'$ that is thermalized with the SM bath in the early Universe\footnote{We remain agnostic on the interaction with the SM that lead to thermalization.
While we note that the addition of a $Z'$ may be subject to other constraints, we do not consider them in this work.} such that its PSD is a Bose-Einstein distribution with temperature~$T$.
The DM particle~$\chi$ interacts with the $Z'$ particle according to the Lagrangian  
\begin{equation}
    \mathcal{L} \supset g_{\chi} Z^{'}_{\mu} \overline{\chi} \gamma^{\mu} \chi + \frac{1}{2} m^2_{Z'}Z^{'}_{\mu}Z^{'\mu} + \bar{\chi}(i \partial - m_{\chi})\chi \, ,
\end{equation}
where $m_\chi$ and $m_{Z'}$ are the DM and $Z'$ masses, respectively, and $g_\chi$ is the coupling constant for the interaction.
The spin-averaged\footnote{We follow the convention in ref.~\cite{kolb:earlyuniversebook} of averaging over initial \textit{and} final states.} scattering amplitude squared for the decay process~$Z' \to \chi \bar{\chi}$ is
\begin{equation}
    \overline{|\mathcal{M}_{Z'\to \chi \overline{\chi}}|}^2 = \frac{g_\chi^2}{24} \left(m_{Z'}^2 + 2m_\chi^2\right) \, .
\end{equation}
The inverse-decay process~$\chi \bar{\chi } \to Z'$ and effects coming from Pauli blocking factors are negligible throughout freeze-in due to the small abundance of~$\chi$ particles (i.e., $f_{\chi} \ll f_{Z'}$).
Under these approximations the collision term for the decay process simplifies to 
\begin{equation}
    \begin{aligned}
        C^{\text{d}}[f_{Z'}] &= \frac{1}{2} \int \frac{3 d^3 p_{Z'}}{(2 \pi)^3 2E_{Z'}} \int \frac{2d^3 p_{\overline{\chi}}}{(2\pi)^3 2E_{\overline{\chi}}} (2\pi)^4 \delta^{(4)} (p_{Z'}-p_{\chi} - p_{\overline{\chi}}) \overline{|\mathcal{M}_{Z'\to \chi \overline{\chi}}|}^2 f_{Z'},
    \end{aligned} 
\label{eq:decay_collision_term}
\end{equation}
where~$\delta^{(n)}$ is an $n$-th dimensional Dirac delta function.


\subsubsection{2-to-2 annihilation}
\label{subsubsec:two_to_two}

For 2-to-2 annihilation, we consider a millicharged DM particle~$\chi$ with mass~$m_\chi > 1 \, \text{MeV}$.
Millicharged DM could arise from a DM hypercharge or from coupling DM to an ultralight dark photon mediator that mixes kinetically with the SM photon; for the purposes of this work, we assume a pure millicharge scenario.

Freeze-in proceeds via electron-positron annihilation, characterized by the Lagrangian
\begin{equation}
    \mathcal{L} \supset e J^{\mu}_{\text{EM}} A_{\mu} + e Q_\chi \overline{\chi}\gamma^{\mu} \chi A_{\mu} + \bar{\chi}(i \partial - m_{\chi})\chi - \frac{1}{4}F^{\mu\nu}F_{\mu \nu} \, ,
\label{eq:mQLagrangian}
\end{equation}
where~$A_\mu$ is the SM photon, $F^{\mu \nu}$ is the electromagnetic field strength, and $J^\mu_{\text{EM}}$ is the SM electromagnetic current.
The spin-averaged amplitude squared of the process~$e_+e_- \to \chi\bar{\chi}$ is
\begin{equation}
    \overline{|\mathcal{M}_{e_{+}e_{-}\to \chi\bar{\chi}}|}^2 = \frac{e^4Q_\chi^2}{2s^2}\left[t^2 + u^2 + 4s\left(m^2_e+m^2_\chi\right)-2\left(m^2_e+m^2_\chi\right)^2 \right],
\end{equation}
where~$s$, $t$, and $u$ are the Mandelstam variables, $m_e$ is the electron/positron mass, and~$Q_\chi e$ is the DM millicharge.
Similar to the decay case, the inverse process~$\chi\bar{\chi} \to e_+e_- $ and Pauli blocking factors are highly suppressed due to the low DM abundance (i.e., $f_{\chi} \ll f_e$).
We estimate the relative suppression in abundance to be
\begin{equation} 
    \frac{f_{\chi}}{{f_e}} \sim \frac{z_{\text{eq}}}{z_{\text{fi}}} \sim 10^{-6} \left(\frac{\text{MeV}}{m_{\chi}}\right) \, ,
\label{eq:abundance}
\end{equation}
where~$z_\mathrm{fi}$ is the redshift at the end of the freeze-in process, $z_\mathrm{eq} \approx 3500$ is the redshift of matter-radiation equality, and we have assumed that DM is nonrelativistic at the end of freeze-in at redshift $z_{\text{fi}}$ and that $f_e(z_{\text{fi}}) \sim f_{\gamma}(z_{\text{fi}})$ for $m_{\chi} \gtrsim 1 \, \text{MeV}$. 
We use a Boltzmann distribution to describe the PSD of the electrons and positrons such that $f_e \approx e^{-E_e/T}$.
The collision term for annihilation simplifies to
\begin{equation}
    \begin{aligned}
        C^{\text{ann}}[f_{e_-},f_{e_+}] &= \frac{1}{2} \int \frac{2 d^3 p_{e_-}}{(2\pi)^3 2 E_{e_-}} \int \frac{2 d^3 p_{e_+}}{(2\pi)^3 2 E_{e_+}} \int \frac{2 d^3 p_{\overline{\chi}}}{(2\pi)^3 2 E_{\overline{\chi}}} \\ 
        &\qquad\qquad\times (2\pi)^4 \delta^{(4)}(p_{e_-}+p_{e_+}-p_\chi-p_{\overline{\chi}}) \overline{|\mathcal{M}_{e_{+}e_{-}\to \chi\bar{\chi}}|}^2 f_{e_-} f_{e_+} .
    \end{aligned}
\label{eq:annihilation_collision_term}
\end{equation}

The Lagrangian in eq.~\eqref{eq:mQLagrangian} also allows for elastic scattering processes~$e^{-} \chi \to e^{-} \chi$ with a scattering amplitude squared of
\begin{equation}
    \overline{|\mathcal{M}_{e_{-}\chi \to e_{-}\chi}|}^2 = \frac{e^4 Q_\chi^2}{2t^2} \left[s^2 + u^2 - 4(m_e^2 + m^2_{\chi})(s+u) + 6(m^2_e + m^2_\chi)^2 \right] \, . 
\end{equation} 
Efficient elastic scattering can establish kinetic equilibrium and alter the DM PSD function.
By writing down the collision term for the elastic scattering process and comparing with eq. \eqref{eq:annihilation_collision_term}, we find that the collision terms of the two processes obey
\begin{equation} 
    C^\mathrm{el} \lesssim C^\mathrm{ann} \frac{\overline{|\mathcal{M}_{e_{-}\chi \to e_{-}\chi}|}^2}{\overline{|\mathcal{M}_{e_{+}e_{-}\to \chi\bar{\chi}}|}^2} \frac{f_{\chi}}{f_e} \, .
\label{eq:2to2_elastic_scattering_estimate}
\end{equation}
At the end of freeze-in, we have
\begin{equation}
    \frac{\overline{|\mathcal{M}_{e_{-}\chi \to e_{-}\chi}|}^2}{\overline{|\mathcal{M}_{e_{+}e_{-}\to \chi\bar{\chi}}|}^2} \approx \frac{s^2}{t^2} \leq \frac{s^2}{m_D^4} \approx 10^4 \, ,
    \label{eq:scattering_amplitudes}
\end{equation}
where~$m^2_D = e^2 T^2 / 3$ is the square of the Debye plasma mass, which accounts for screening of the electromagnetic field in a charged, relativistic plasma~\cite{Blaizot:1995kg}.
Therefore, during the DM creation process,
\begin{equation}
    C^\mathrm{el}/C^\mathrm{ann} \lesssim 0.01 \left(\frac{\text{MeV}}{m_{\chi}}\right)  \left(\frac{s^2/m^4_D}{10^4} \right) \, ,
\label{eq:suppression}
\end{equation}
which indicates that the evolution of the PSD function in eq.~\eqref{eq:boltzmann_equation} is dominated by the s-channel annihilation collision term in eq.~\eqref{eq:annihilation_collision_term}. 
Thus, we estimate that elastic processes are subdominant to the leading collision term due to annihilations~$C^{\text{ann}}$ for the mass range we consider ($m_{\chi} \geq 1 \, \text{MeV})$ and can be neglected when solving eq.~\eqref{eq:boltzmann_equation}.

After freeze-in, when DM is nonrelativistic, we expect the rate of momentum transfer between DM and the baryon fluid to be sufficiently small, such that~$f_{\chi}$ remains unaltered. 
Based on CMB studies of millicharged DM scattering~\cite{PhysRevD.98.123506, Buen-Abad:2021mvc, nguyen:dmescatteringcmb}, for $m_\chi \gtrsim 1~\mathrm{MeV}$, the momentum-transfer rate coefficient between DM and electrons/protons is $\lesssim \mathcal{O}(0.01)$ of the Hubble expansion rate pre-recombination for $Q_{\chi} \gtrsim \mathcal{O}(10^{-9})$.
Since at least an order of magnitude smaller millicharge $Q_\chi e$ is required to reproduce the observed DM abundance, we expect elastic scattering to have a negligible impact on the DM velocity distribution.
In summary, we calculate $f_{\chi}$ from the annihilation collision term $C^{\text{ann}}$ throughout freeze-in and assume that it remains unaltered afterwards.


\subsection{Background evolution}
\label{subsec:background_cosmology}

The Boltzmann equation~\eqref{eq:boltzmann_equation} for the background PSD $\bar{f}_{\chi}$ simplifies to 
\begin{equation}
    \frac{\partial\bar{f}_\chi}{\partial\tau} = \frac{a^2}{\epsilon} \bar{C}\, ,
\label{eq:boltzmann_equation_background}
\end{equation}
where~$\epsilon \equiv aE = \sqrt{q^2+m^2_\chi a^2}$ is the comoving energy, and $\vec{q} \equiv a\vec{p}$ is the comoving momentum.
The background collision terms in eq.~\eqref{eq:decay_collision_term} and~\eqref{eq:annihilation_collision_term} depend only on the homogeneous background temperature~$\bar{T}(\tau)$, which coincides with the photon temperature in the early Universe.
Therefore, the background collision term in the decay case is
\begin{equation}
    \begin{aligned}
        C^{\text{d}}[f_{Z'}] &= \frac{g^2_{\chi}}{64 \pi p_{\chi}} (m^2_{Z'} + 2 m^2_{\chi}) \int_{E_{\text{min}}}^{E_{\text{max}}} dE \frac{1}{e^{E/\bar{T}}-1} \, ,
    \end{aligned} 
\end{equation}
where
\begin{equation}
    E_\mathrm{min} = \frac{m^2_{Z'}}{2m^2_\chi} \left( E_\chi - p_\chi\sqrt{1-4m^2_\chi/m^2_{Z'}} \right),\quad E_\mathrm{max} = \frac{m^2_{Z'}}{2m^2_\chi} \left( E_\chi + p_\chi\sqrt{1-4m^2_\chi/m^2_{Z'}} \right) \, ,
\end{equation}
and in the annihilation case is
\begin{equation}
    \begin{aligned}
        C^{\text{ann}}[f_{e_-},f_{e_+}] &= \frac{2\bar{T} \alpha^2 Q^2 }{3\pi p_\chi } \int_{s_{\text{min}}}^{\infty} \frac{ds}{s^2} e^{-\frac{E_{\chi}s }{2 m_{\chi}\bar{T}}} \sinh{\left(\frac{p_{\chi}\sqrt{s(s-4m^2_{\chi})}}{2 m^2_{\chi}\bar{T}}\right) \sqrt{1-\frac{4m^2_e}{s}} (2 m^2_e + s)(2m^2_{\chi} + s)} \, ,
    \end{aligned}
\end{equation}
where~$s_{\text{min}} = \text{max}\left[ 4 m^2_e, 2m_{\chi}(E_{\chi}+m_{\chi}) \right]$.

Macroscopic quantities like the DM number density, energy density, and pressure are obtained from the PSD $\bar{f}_{\chi}$ by
\begin{align}
    \bar{n}_{\chi}(\tau) = \int \frac{2d^3p}{(2 \pi)^3} \bar{f}_{\chi}(\vec{p},\tau) \, , \label{eq:n} \\
    \bar{\rho}_{\chi}(\tau) = \int \frac{2d^3p}{(2 \pi)^3} E \bar{f}_{\chi}(\vec{p},\tau)    \, , \label{eq:rho} \\
    \bar{p}_{\chi}(\tau) = \int \frac{2d^3p}{(2 \pi)^3} \frac{p^2}{3E} \bar{f}_{\chi}(\vec{p},\tau) \, . \label{eq:P}
\end{align}
Integrating eq.~\eqref{eq:boltzmann_equation_background} recovers both the number density and energy density conservation equations
\begin{align}
    \frac{d\bar{n}_{\chi}}{d\tau} &+ 3 \mathcal{H}\bar{n}_{\chi} = \frac{1}{a^2} \int \frac{2d^3q}{(2 \pi)^3} \frac{\bar{C}}{\epsilon} \, \label{eq:numberdensity_continuity}, \\
    \frac{d\bar{\rho}_{\chi}}{d\tau} &+ 3 \mathcal{H}\bar{\rho}_{\chi}(1 + w_{\chi}) = a \bar{\mathcal{Q}}  \, \label{eq:energydensity_continuity},
\end{align}
where $\mathcal{H} = a^{-1}\frac{da}{d\tau}$ is the conformal Hubble expansion rate, $w_{\chi} = \bar{p}_{\chi}/\bar{\rho}_{\chi}$ is the DM equation of state, and~$\bar{\mathcal{Q}}$ is the energy exchange term defined as
\begin{equation}
    \bar{\mathcal{Q}} \equiv \frac{1}{a^3} \int \frac{2d^3q}{(2 \pi)^3} \bar{C} \, .
\end{equation}
The last macroscopic DM property relevant for our analysis is the adiabatic sound speed~$c^2_{a,\chi}=\bar{p}'_\chi/\bar{\rho}'_\chi$.
Freeze-in calculations of DM models~\cite{Hall:2009bx, Yaguna:2011qn, Chu:2011be} often entail solving eq.~\eqref{eq:numberdensity_continuity} or eq.~\eqref{eq:energydensity_continuity} to relate the predicted amount of DM for a given interaction and coupling strength to the observed relic abundance of DM.
However, in this work, we solve for the DM PSD directly. 
Predictions for the shape of the nonthermal phase space are only accessible when solving eq.~\eqref{eq:boltzmann_equation_background} directly.
Moreover, the evolution of macroscopic DM fluid properties, such as the equation of state, can be only be derived from first principles using a PSD approach.

\begin{figure}[t]
    \centerline{
    \includegraphics[width=\columnwidth]{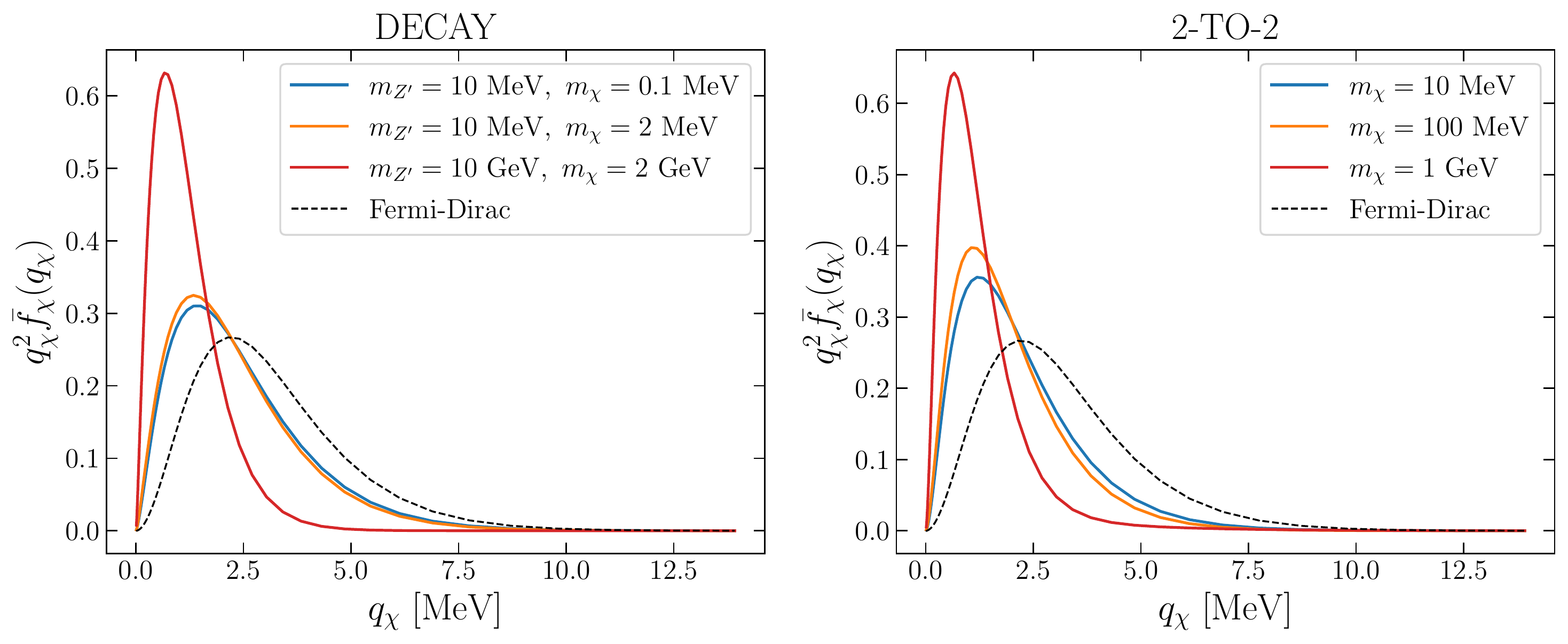}}
    \caption{Background PSD of freeze-in DM for the decay (\textit{left panel}) and 2-to-2 annihilation (\textit{right panel}) scenarios, as a function of comoving momentum.
    The normalization for each PSD is set such that it integrates to 1, and the scale factor is normalized to be unity at~$T=1\ \mathrm{MeV}$.
    The dashed, black line represents the Fermi-Dirac distribution.}
    \label{fig:bg_psd}
\end{figure}

We solve eq.~\eqref{eq:boltzmann_equation_background} in the interval~$[\tau_\mathrm{ini},\tau_\mathrm{fin}]$ for the collision terms in eq.~\eqref{eq:decay_collision_term} and eq.~\eqref{eq:annihilation_collision_term}, corresponding to our two models of interest.
We choose the initial time such that~$\bar{T}(\tau_\mathrm{ini}) \gg \bar{T}(z_\mathrm{fi})$, while the final time~$\tau_\mathrm{fin}$ is chosen such that~$\bar{\mathcal{Q}}(\tau_{\text{fin}}) = 0$ and~$m_{\chi} \gg p_{\chi}(\tau_{\text{fin}})$, i.e., well after DM has transitioned to become nonrelativitistic.
We use 60 bins of discrete comoving momenta spanning from~$q_\chi =1.4\times 10^{-2}\ \mathrm{MeV}$ to~$14\ \mathrm{MeV}$ for all DM masses considered, normalizing the scale factor to unity at~$T=1\ \mathrm{MeV}$.
We take~$\bar{f}_{\chi}(\tau_{\mathrm{ini}}) = 0$, noting that the two classes of models we consider have an attractor solution.
In other words, irrespective of the exact initial condition for the DM PSD, the IR-dominated dynamical evolution leads to the same final PSD; thus, we are not sensitive to the exact choice of $\bar{f}_{\chi}(\tau_{\mathrm{ini}})$.
figure~\ref{fig:bg_psd} shows the resulting PSDs for various choices of model parameters, along with the Fermi-Dirac distribution for comparison.
The peak of the PSD for more massive DM candidates is shifted towards lower momenta with respect to the PSD of lighter ones.
DM is created relativistic, with a typical momentum of order~$p_\chi\sim \bar{T}(z_\mathrm{fi}) = \bar{T}_\mathrm{fi}$.
This shift in the PSD peak between heavier and lighter species is approximately~$\left[g_{\ast}(\bar{T}^\mathrm{lighter}_\mathrm{fi})/g_{\ast}(\bar{T}^\mathrm{heavier}_\mathrm{fi})\right]^{1/3}<1$.

\begin{figure}[t]
    \centerline{
    \includegraphics[width=\columnwidth]{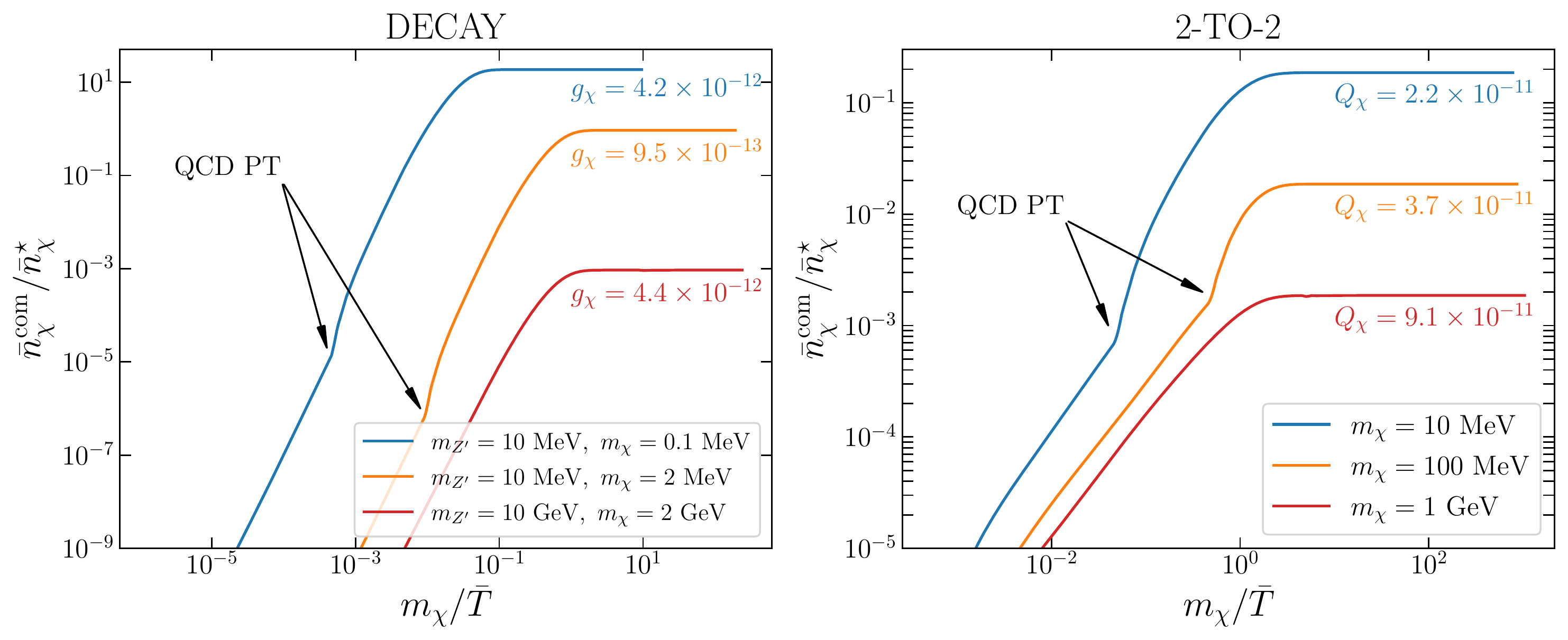}}
    \caption{Background decay (\textit{left panel}) and freeze-in millicharged DM (\textit{right panel}) comoving number density.
    Numbers are normalized by~$\bar{n}^\star_\chi=10^{70}\ \mathrm{MeV\ Mpc^{-3}}$. 
    The kinks appear at the beginning of the QCD phase transition (PT).
    Couplings are reported below each corresponding line.}
    \label{fig:bg_number_density}
\end{figure}

In figure~\ref{fig:bg_number_density}, we show the evolution of the background comoving number density~$\bar{n}^\mathrm{com}_\chi = \bar{n}_\chi a^3$ as a function of $m_\chi / \bar{T}$.
Note that temperature scales with $a$ according to entropy conservation: $\bar{T}^{-1} \sim g_\ast (\bar{T})^{1/3} a$.\footnote{Since we consider freeze-in processes that complete before neutrino decoupling, $g_\ast$ as defined in eq.~\eqref{eq:energy-density-bath} coincides with $g_{\ast S}$, the effective number of relativistic degrees of freedom associated with entropy density.}
Therefore, $\bar{T}$ decreases less slowly as a function of $a$ when $g_\ast$ decreases, particularly after the QCD phase transition at temperature $\bar{T} = 200~\mathrm{MeV}$, and this effect is noted for the curves in figure~\ref{fig:bg_number_density}.
We fix the coupling constants~$g_{\chi}$ and~$Q_{\chi}$ by matching the comoving number density abundance~$\bar{n}^\mathrm{com}_\chi=\bar{n}_\chi a^3$ such that freeze-in produces all of the observed dark matter abundance today, requiring~$\bar{n}^\mathrm{com}_\chi(\tau_{\text{fin}}) = \bar{n}_{\chi}(\tau_{\text{fin}})a(\tau_{\text{fin}})^3 = \bar{n}_{\chi,0}$.

While~$\bar{f}(\tau_\mathrm{fin})$ does not depend on~$\bar{f}(\tau_\mathrm{ini})$, the required value of the coupling constant needed to produce the observed DM abundance does depend on whether a fraction of DM was created during reheating, yielding an initial DM abundance~$\bar{n}^\mathrm{ini}_\chi$.
Since the freeze-in DM abundance scales approximately with the coupling constant squared, we have that~$Q^{(f)}_{\chi}/Q^{(0)}_{\chi}\approx (1-\mathcal{F}_\mathrm{ini})^{1/2}$ for the case of 2-to-2 annihilation, where~$Q^{(f)}_{\chi}$ and~$Q^{(0)}_{\chi}$ correspond to the couplings for which~$\bar{f}_\chi\neq 0$ and~$\bar{f}_\chi= 0$ at initial time $\tau_\mathrm{ini}$, respectively.
The parameter~$\mathcal{F}_\mathrm{ini} = \bar{n}^\mathrm{com}_\chi(\tau_\mathrm{ini}) / \bar{n}^\mathrm{com}_\chi(\tau_\mathrm{fin})$ quantifies the initial comoving abundance of DM. 


\subsection{Perturbative evolution}
\label{subsec:perturbative_cosmology}

To solve for the spatial fluctuations of freeze-in DM, we choose to work in synchronous gauge, in which the line element squared
\begin{equation}
   ds^2 = a^2(\tau) \left[-d\tau +(\delta_{ij} + h_{ij}) dx^i dx^j \right]
\end{equation}
contains the metric perturbation
\begin{equation}
   h_{ij}(\vec{x},\tau) = \int d^3 k e^{i \vec{k} \cdot \vec{x}} \left[\frac{k_i k_j}{k^2} h(\vec{k},\tau) +\left(\frac{k_i k_j}{k^2}-\frac{1}{3}\delta_{ij}\right)6\eta(\vec{k},\tau) \right] \, . 
\end{equation}
The Boltzmann equation in Fourier space for the perturbed PSD~$\widehat{\delta f}_\chi$ reads \cite{ma:perturbationtheory}
\begin{equation}
    \frac{\partial \widehat{\delta f}_\chi}{\partial \tau} +  \frac{i}{\epsilon}(\vec{k}\cdot\vec{q})\widehat{\delta f}_\chi + \frac{d \bar{f}_\chi}{d\log q}\left[\frac{d\eta}{d\tau} - \frac{(\vec{k}\cdot\vec{q})^2}{2k^2q^2} \left(\frac{dh}{d\tau} +\frac{d\eta}{d\tau} \right)  \right] = \frac{a^2}{\epsilon} \widehat{\delta C} \, ,
\label{eq:boltzmann_equation_perturbed}
\end{equation}
where $\widehat{\delta C}$ is the perturbed collision term in Fourier space determined by
\begin{equation}
    \begin{aligned}
        \widehat{\delta C}^\mathrm{d}(\vec{k}, \vec{p},\tau) &= \frac{1}{2} \int \frac{3d^3p_{Z'}}{(2\pi)^32E_{Z'}} \int \frac{2d^3p_{\overline{\chi}}}{(2\pi)^3 2E_{\overline{\chi}}}      \delta^{(4)}(p_{Z'}-p_{\chi}-p_{\overline{\chi}}) \overline{|\mathcal{M}_{Z'\to \chi \overline{\chi}}|}^2 \widehat{\delta f}_{Z'}( \vec{k}, \vec{p}_{Z'}, \tau) \, ,
    \end{aligned}
\end{equation}
\begin{equation}
    \begin{aligned}
        \widehat{\delta C}^{\text{ann}}( \vec{k}, \vec{p},\tau) &= \frac{1}{2} \int \frac{2d^3 p_{e_-}}{(2\pi)^32E_{e_-}} \int \frac{2d^3 p_{e_+}}{(2\pi)^3 2E_{e_+}}  \int \frac{2d^3 p_{\overline{\chi}}}{(2\pi)^3 2E_{\overline{\chi}}} (2\pi)^4 \delta^{(4)}(p_{e_-}+p_{e_+}-p_{\chi}-p_{\overline{\chi}}) \\ 
        &\qquad\qquad \times \overline{|\mathcal{M}_{e_{+}e_{-}\to \chi\bar{\chi}}|}^2 \left[\widehat{\delta f}_{e_-}(\vec{k}, \vec{p}_{e_-},\tau) \bar{f}_{e_+}( \vec{p}_{e_+},\tau) + \bar{f}_{e_-}( \vec{p}_{e_-},\tau) \widehat{\delta f}_{e_+}(\vec{k}, \vec{p}_{e_+},\tau) \right]\, ,
    \end{aligned}
\end{equation}
for the decay and annihilation freeze-in models, respectively.
For a fixed wavenumber $k$, eq.~\eqref{eq:boltzmann_equation_perturbed} depends on the conformal time~$\tau$, the magnitude of the comoving momenta $q$, and $\hat{\vec{k}}\cdot \hat{ \vec{q}} = \vec{k}\cdot \vec{q}/(kq)$. 
We expand the perturbed PSD as a Legendre series:
\begin{equation}
    \widehat{\delta f}_{\chi}(\vec{k}, \vec{p}, \tau) = \sum_\ell (-i)^\ell (2\ell+1) \widehat{\delta f}_{\chi,\ell}(\vec{k}, \vec{p},\tau) P_\ell (\hat{\vec{k}}\cdot\hat{\vec{p}}) \, , 
\end{equation}
where~$P_\ell$ are Legendre polynomials. We similarly expand the perturbed collision term.
The resulting Boltzmann hierarchy is
\begin{equation}
    \begin{aligned}
        \frac{\partial \widehat{\delta f}_{\chi,0}}{\partial \tau} &= -\frac{qk}{\epsilon}\widehat{\delta f}_{\chi,1} + \frac{1}{6}\frac{dh}{d\tau}\frac{d \bar{f}_\chi}{d\log q} + \frac{a^2}{\epsilon} \widehat{\delta C}_{0}, \\
        \frac{\partial \widehat{\delta f}_{\chi,1}}{\partial \tau} &= \frac{qk}{3\epsilon}\left[\widehat{\delta f}_{\chi,0} - 2\widehat{\delta f}_{\chi,2} \right] + \frac{a^2}{\epsilon} \widehat{\delta C}_{1}, \\
        \frac{\partial \widehat{\delta f}_{\chi,2}}{\partial \tau} &= \frac{qk}{5\epsilon}\left[2 \widehat{\delta f}_{\chi,1} - 3\widehat{\delta f}_{\chi,3} \right] - \left(\frac{1}{15}\frac{dh}{d\tau}+\frac{2}{5}\frac{d\eta}{d\tau}\right)\frac{d\bar{f}_\chi}{d\log q} + \frac{a^2}{\epsilon} \widehat{\delta C}_{2}, \\
        \frac{\partial \widehat{\delta f}_{\chi,\ell}}{\partial \tau} &= \frac{qk}{(2\ell+1)\epsilon}\left[\ell \widehat{\delta f}_{\chi,\ell-1} - (\ell+1)\widehat{\delta f}_{\chi,\ell+1} \right] + \frac{a^2}{\epsilon} \widehat{\delta C}_{\ell}, \qquad\quad \mathrm{for}\ \ell \geq 3.
    \end{aligned}
\label{eq:boltzmann_hierarchy_synchronous_gauge}
\end{equation}
We focus on deriving DM perturbations on the large scales probed by CMB experiments, so we solve the Boltzmann hierarchy well before horizon crossing (i.e., $k\tau \ll 1$).
In this limit, $\widehat{\delta f}_{\chi,\ell\geq 3} \simeq (k\tau)^{\ell} \approx 0 $, which allows us to restrict our analysis to multipoles~$\ell=0,1,2$.

To calculate the multipole moments of the collision term, we expand the perturbed PSD in terms of temperature fluctuations $\delta T(\vec{x},\hat{\vec{p}},\tau)$ in the SM bath.
For the decay case, the PSD for $Z'$ is 
\begin{equation}
    f_{Z'}(\vec{x},\vec{p},\tau) \approx  \left(e^{E/\bar{T}}-1\right)^{-1} + \frac{E }{\bar{T}^2}\frac{e^{E/\bar{T}}}{(e^{E/\bar{T}}-1)^2} \delta T(\vec{x},\hat{\vec{p}},\tau)  \, ,
\end{equation}
assuming that rapid elastic processes in the thermal bath suppress any dependence on the absolute value of the momentum~\cite{bhatnagar:rta, hannestad:rta, oldengott:rta, eganaugrinovic:rta}.
With this simplification we find
\begin{equation}
    \begin{aligned}
        \widehat{\delta C}^\mathrm{d}_0 &= \frac{g_\chi^2\left(m_{Z'}^2 + 2m_\chi^2\right)\widehat{\delta T}_0} {64 \pi  p_\chi} \left[ \frac{E_\mathrm{min}/\bar{T}}{e^{E_\mathrm{min}/\bar{T}}-1} -\frac{E_\mathrm{max}/\bar{T}}{e^{E_\mathrm{max}/\bar{T}}-1} + \log\frac{1-e^{-E_\mathrm{max}/\bar{T}}}{1-e^{-E_\mathrm{min}/\bar{T}}} \right] \, ,\\
        \widehat{\delta C}^\mathrm{d}_1 &= \frac{g_\chi^2\left(m_{Z'}^2 + 2m_\chi^2\right)\widehat{\delta T}_1}{64 \pi  p_\chi} \int^{E_\mathrm{max}}_{E_\mathrm{min}} \frac{dE}{\bar{T}} \frac{E}{\bar{T}} \frac{e^{E/\bar{T}}}{(e^{E/\bar{T}}-1)^2} \frac{2EE_\chi-m^2_{Z'}}{2p_\chi\sqrt{E^2-m^2_{Z'}}} \, ,
    \end{aligned}  
\label{eq:decay_perturbed_collision_term}
\end{equation}
where the temperature fluctuation in Fourier space has also been expanded in a Legendre series as
\begin{equation}
    \widehat{\delta T}(\vec{k},\hat{\vec{p}},\tau) = \sum_\ell (-i)^\ell (2\ell+1) \widehat{\delta T}_{\ell}(k,\tau) P_\ell (\hat{\vec{k}}\cdot\hat{\vec{p}}) \, . 
\end{equation}
Anisotropic stress vanishes for a species in thermal equilibrium; thus, $\widehat{\delta T}_2 = 0$ and $\widehat{\delta C}_2 \propto \widehat{\delta T}_2 = 0$. 
Therefore, eq.~\eqref{eq:boltzmann_hierarchy_synchronous_gauge} together with eq.~\eqref{eq:decay_perturbed_collision_term} specify the full Boltzmann hierarchy.
In a similar fashion, for millicharged DM we expand the electron/positron PSD as
\begin{equation}
    f_e (\vec{x}, \vec{p}, \tau) \approx e^{-E/\bar{T}} + \frac{E}{\bar{T}^2} e^{-E/\bar{T}} \delta T(\vec{x}, \hat{\vec{p}}, \tau),
\end{equation}
to compute the multipole moments of the perturbed collision term and find 
\begin{equation}
    \begin{aligned}
        \widehat{\delta C}^{\text{ann}}_0 &= \frac{Q^2_\chi e^4  \widehat{\delta T }_0}{12(2\pi)^3 p_\chi} \int_{s_\mathrm{min}} ds \left[e^{-E_\mathrm{min}/\bar{T}} \left(\frac{E_\mathrm{min}}{\bar{T}}+1\right) - e^{-E_\mathrm{max}/\bar{T}} \left(\frac{E_\mathrm{max}}{\bar{T}}+1\right)\right] \\
        &\qquad\qquad\qquad\qquad\qquad\qquad \times \sqrt{1-\frac{4m_e^2}{s}} \left[1+\frac{2m_e^2}{s}\right] \left[1+\frac{2m_\chi^2}{s}\right]\, , \\
        \widehat{\delta C}^\mathrm{ann}_1 &= \frac{Q^2_\chi e^4  \widehat{\delta T}_1}{12(2\pi)^3 p_\chi} \int_{s_\mathrm{min}} ds \sqrt{1-\frac{4m_e^2}{s}} \left[1+\frac{2m_e^2}{s}\right] \left[1+\frac{2m_\chi^2}{s}\right]  \\
        &\qquad\qquad\qquad \times \left[e^{-E_\mathrm{min}/\bar{T}} \left(\frac{E_\mathrm{min}}{\bar{T}}+1-\frac{s}{2\bar{T}E_\chi}\right) - e^{-E_\mathrm{max}/\bar{T}} \left(\frac{E_\mathrm{max}}{\bar{T}}+1-\frac{s}{2\bar{T}E_\chi}\right)\right].
    \end{aligned}
\end{equation}
We report additional details of the calculation in appendix~\ref{app:perturbed_collision_terms}.

We obtain the perturbed macroscopic fluid quantities---such as the energy density fluctuation $\delta\rho_\chi$, pressure perturbation $p_\chi$, velocity divergence $\theta_\chi$, and anisotropic stress $\sigma_\chi$---by integrating the Boltzmann hierarchy in eqs.~\eqref{eq:boltzmann_hierarchy_synchronous_gauge} (see appendix~\ref{app:fluid_perturbations} for details)~\cite{ma:perturbationtheory, shoji:highermoments, lesgourgues:classncdm}:
\begin{equation}
  \begin{aligned}
    \delta\rho_\chi &= \delta_{\chi} \bar{\rho}_{\chi} = a^{-4} \int \frac{2d^3q}{(2\pi)^3} \epsilon \widehat{\delta f}_{\chi,0}
    \, , &
    (\bar{\rho}_\chi+\bar{p}_\chi) \theta_\chi &= ka^{-4} \int \frac{2d^3q}{(2\pi)^3} q \widehat{\delta f}_{\chi,1}
    \, , \\
    \delta p_\chi &= a^{-4} \int \frac{2d^3q}{(2\pi)^3} \frac{q^2}{3\epsilon} \widehat{\delta f}_{\chi,0}
    \, , &
    (\bar{\rho}_\chi+\bar{p}_\chi) \sigma_\chi &= a^{-4} \int \frac{2d^3q}{(2\pi)^3} \frac{2q^2}{3\epsilon} \widehat{\delta f}_{\chi,2}
    \, .
  \end{aligned}
\end{equation}
Since DM is created relativistic, the fluid equations during freeze-in can be approximated as
\begin{equation}
    \begin{aligned}
        &\frac{d\delta\rho_\chi}{d\tau} + 3\mathcal{H}\delta\rho_\chi+3\mathcal{H} (\delta \rho_{\chi} + \delta p_{\chi}) + \bar{\rho}_\chi(1+w_{\chi})
        \left(\theta_\chi + \frac{1}{2}\frac{dh}{d\tau}\right) = a \widehat{\delta \mathcal{Q}}_{0}\, , \\
        &\frac{d\left[(\bar{\rho}_\chi+\bar{p}_\chi)\theta_\chi\right]}{d\tau} + 4\mathcal{H}\bar{\rho}_\chi (1+w_{\chi})
        \theta_\chi - k^2\delta p_\chi + k^2\bar{\rho}_\chi(1+w_{\chi})\sigma_\chi = a \widehat{\delta \mathcal{Q}}_{1}\, , \\
        &\frac{d\left[(\bar{\rho}_\chi+\bar{p}_\chi)\sigma_\chi\right]}{d\tau} + 4\mathcal{H}\bar{\rho}_\chi(1+w_{\chi})\sigma_{\chi} - \frac{4}{15}\bar{\rho}_\chi(1+w_{\chi}){\theta}_\chi \\&\qquad\qquad\qquad\qquad\qquad\qquad\qquad\qquad\qquad\qquad\qquad - \left(\frac{8}{15}\frac{dh}{d\tau}+\frac{16}{5}\frac{d\eta}{d\tau}\right)\bar{p}_\chi = 0 \, ,
    \end{aligned}
    \label{eq:perturbed_fluid_equations1}
\end{equation}    
where the source terms on the right-hand side of the equations represent the energy and momentum exchange, given by
\begin{equation}
    \widehat{\delta \mathcal{Q}}_{0} \equiv  a^{-3} \int \frac{2d^3q}{(2\pi)^3} \widehat{\delta C}_{0}
    \qquad \textrm{and} \qquad
    \widehat{\delta \mathcal{Q}}_{1} \equiv ka^{-3} \int \frac{2d^3q}{(2\pi)^3} \frac{q}{\epsilon} \widehat{\delta C}_{1} \, ,
\end{equation}
respectively.
We obtain $\widehat{\delta \mathcal{Q}}_{0}$ and~$\widehat{\delta \mathcal{Q}}_{1}$ numerically by directly integrating the collision terms.
We also derive an analytic approximation of $\widehat{\delta \mathcal{Q}}_{0}$ and~$\widehat{\delta \mathcal{Q}}_{1}$ (see appendix~\ref{app:perturbed_collision_terms} for details) for both models of interest in the regime where~$Z'$ and~$e_+/e_-$ are relativistic:
\begin{align}
    \widehat{\delta \mathcal{Q}}^\mathrm{d}_{0} &\approx \mathcal{C}_{0,\mathrm{d}} \frac{g^2_\chi (m^2_{Z'}+2m^2_\chi) \widehat{\delta T}_0}{16 (2\pi)^3} \sqrt{1-\frac{4m^2_\chi}{m^2_{Z'}}} \int dE \sqrt{E^2-m^2_{Z'}} \left(\frac{E}{\bar{T}}\right)^{2} \frac{e^{E/\bar{T}}}{(e^{E/\bar{T}}-1)^2} \, , \label{eq:perturbative_energyexchange_decay} \\
    \widehat{\delta \mathcal{Q}}^\mathrm{d}_{1} &\approx \mathcal{C}_{1,\mathrm{d}} \frac{g^2_\chi (m^2_{Z'}+2m^2_\chi) k \widehat{\delta T}_1}{16 (2\pi)^3} \sqrt{1-\frac{4m^2_\chi}{m^2_{Z'}}} \int dE \left(E^2-m^2_{Z'}\right)^{3/2} \left(\frac{E}{\bar{T}}\right)^2 \frac{e^{E/\bar{T}}}{(e^{E/\bar{T}}-1)^2} \, , \label{eq:perturbative_momentumexchange_decay} \\
    \widehat{\delta \mathcal{Q}}^{\text{ann}}_{0} &\approx \mathcal{C}_{0,\mathrm{ann}}\frac{e^4 Q^2_\chi \widehat{\delta T}_0}{6 (2\pi)^5} \int ds \sqrt{1-\frac{4m_\chi^2}{s}} \sqrt{1-\frac{4m_e^2}{s}} \left(1+\frac{2m_e^2}{s}\right) \left(1+\frac{2m_\chi^2}{s}\right) \nonumber \\
    &\qquad\qquad\qquad \times \left(6+6\frac{\sqrt{s}}{\bar{T}}+3\frac{s}{\bar{T}^2}+\frac{s^{3/2}}{\bar{T}^3}\right) e^{-\sqrt{s}/\bar{T}}  \, , \label{eq:perturbative_energyexchange_2to2} \\
    \widehat{\delta \mathcal{Q}}^{\text{ann}}_{1} &\approx \mathcal{C}_{1,\mathrm{ann}} \frac{e^4 Q^2_\chi k \widehat{\delta T}_1}{6 (2\pi)^5} \int ds \sqrt{1-\frac{4m_\chi^2}{s}} \sqrt{1-\frac{4m_e^2}{s}} \left(1+\frac{2m_e^2}{s}\right) \left(1+\frac{2m_\chi^2}{s}\right) 3 s K_2(\sqrt{s}/\bar{T}) \, , \label{eq:perturbative_momentumexchange_2to2}
\end{align}
where $K_2(x)$ are modified Bessel functions of the second kind, and the temperature multipoles are related to the overdensity and velocity divergence of the thermal species as~$\widehat{\delta T}_0=\bar{T}\delta_{Z^\prime}/4$ ($\widehat{\delta T}_0=\bar{T}\delta_{e}/4$) and~$\widehat{\delta T}_1=\bar{T}\theta_{Z^\prime}/(3k)$ ($\widehat{\delta T}_1=\bar{T}\theta_{e}/(3k)$) for the decay (annihilation) case. 
To align with the exact numerical results of integrating the collision terms, we introduce the calibration constants~$\mathcal{C}_{0\mathrm{d}},\mathcal{C}_{1\mathrm{d}}\approx 2$ and $\mathcal{C}_{0,\mathrm{ann}},\mathcal{C}_{1,\mathrm{ann}} \approx 1.5$.
As these calibration constants are $\mathcal{O}(1)$, the analytic approximations are broadly consistent with the numerical result; thus, we surmise the approximations would be applicable to other models as well.


\subsection{Initial conditions for DM freeze-in perturbations}
\label{subsec:initial_conditions_perturbations}

The last step required to solve the Boltzmann hierarchy is specifying the initial conditions for the perturbed PSD. 
The most general solution of the perturbed Boltzmann equation for a combination of multiple fluids is a linear combination of different modes, which evolve independently. 
The observable modes for a two-component system, consisting of the SM bath and DM, are the adiabatic and DM isocurvature modes.
The adiabatic mode corresponds to the energy density perturbations of the DM being in phase with the energy density perturbations of the SM bath, such that a local overdensity in the SM bath corresponds to an overdensity in the DM. 
The DM isocurvature mode describes the situation in which a local overdensity in the SM bath corresponds to an underdensity in the DM, such that the total energy density is homogenous. 
In principle, there is also a velocity isocurvature mode for a relativistic species; however, such a mode rapidly decays as DM becomes nonrelativistic. 

The classification into adiabatic and isocurvature modes is well-defined on super-horizon scales ($k \tau  \ll 1 $), where they are characterized in terms of a regular (constant or growing in the limit~$k \tau \to 0$) or singular (decaying in the limit~$k \tau \to 0$) modes, when properly accounting for gauge modes~\cite{bucher:initialconditions}. 
We focus on the evolution of perturbations on cosmologically relevant scales~$k\in [10^{-4},10^{-1}]\ \mathrm{Mpc}^{-1}$, which remain in the super-horizon regime for the entire DM-formation epoch (i.e., $k\tau_\mathrm{ini}\ll k\tau_\mathrm{fin}\ll 1$).
Our aim is to evolve the perturbed DM PSD throughout freeze-in for both physical modes.
To find the initial conditions at the beginning of freeze-in for the multipoles in each mode, we consider
\begin{equation}
    \begin{aligned}
        \widehat{\delta f}_{\chi,0} = - \frac{\delta_\chi}{4} \frac{d\bar{f}_\chi}{d\log q}, \qquad
        \widehat{\delta f}_{\chi,1} = - \frac{\epsilon\theta_\chi}{3qk} \frac{d\bar{f}_\chi}{d\log q}, \qquad
        \widehat{\delta f}_{\chi,2} = - \frac{\sigma_\chi}{2} \frac{d\bar{f}_\chi}{d\log q} \, . 
    \end{aligned}
\label{eq:perturbed_PSD_initial_considtions}
\end{equation}
where the attractor solution for~$\delta^{\text{ad/iso}}_\chi,\ \theta^{\text{ad/iso}}_\chi, \sigma^{\text{ad/iso}}_\chi$ at initial time~$\tau_{\text{ini}}$ is found using the iterative procedure of ref.~\cite{doran:initialconditions}. 
In particular, we relate the task of finding initial conditions and observable modes to the language of eigenvectors and eigenvalues of the system of perturbed Einstein equations for the metric variables ($h,\eta$) and conservation laws for the perturbed variables describing the radiation ($\delta_\mathrm{r}, \theta_\mathrm{r}$) and DM ($\delta_\chi, \theta_\chi, \sigma_\chi$) sectors:
\begin{equation}
    \frac{d\vec{U}}{d\log (k\tau)} = A(k \tau)\vec{U},
\label{eq:modes_linear_system}
\end{equation}
where~$\vec{U}=\left(\eta,h,\delta_\mathrm{r},\theta_\mathrm{r},\delta_\chi,\theta_\chi,\sigma_\chi\right)$ and~$A$ is a matrix encoding the evolution equations.

The general solution of eq.~\eqref{eq:modes_linear_system} reads
\begin{equation}
    \vec{U}(k\tau) = \sum_{j} C_j (k \tau/k \tau_\star)^{\lambda_j} \vec{U}^{(j)},
\end{equation}
where the index~$j$ labels the different modes, $C_j$ are constants, $\lambda_j$ are the eigenvalues of the modes, and~$\tau_\star$ is a reference time scale in which we assess the relative importance of different modes.
The matrix~$A$ is expanded in powers of~$k\tau$, i.e., $A(k\tau)=\sum_{j=0} A_j (k\tau)^j$, where~$A_j$ are constant matrices. 
Similarly, we also expand the vector~$\vec{U}^{(j)} = \sum_{i=0} \vec{U}^{(j)}_i (k\tau)^i$, where~$i$ labels the order of the expansion in~$k\tau$.
Thus, $\lambda_j$ and~$U^{(j)}_0$ are the eigenvalues and associated eigenvectors for~$A_0$.
The eigenvalue determines whether the mode is regular ($\lambda_j \geq 0$) or singular ($\lambda_j<0$).

We identify the adiabatic mode as the one with constant~$\eta$ in the limit~$k\tau\to 0$ (i.e., eigenvalue~$\lambda_{\text{ad}} = 0$).
The amplitude of this mode is the curvature perturbation~$\mathcal{R}$.
For the DM attractor solution in the adiabatic mode, we find 
\begin{align}
    h^\mathrm{ad} &= \frac{1}{2}(k\tau)^2\mathcal{R}, \\
    \eta^\mathrm{ad} &= 1-\frac{1}{36}(k\tau)^2\mathcal{R}, \\
    \delta^\mathrm{ad}_\mathrm{r} &= -\frac{1}{3}(k\tau)^2\mathcal{R}, \\
    \theta^\mathrm{ad}_\mathrm{r} &= -\frac{1}{36}k(k\tau)^3\mathcal{R}, \\
    \delta^\mathrm{ad}_\chi &= -\frac{1}{3} \frac{1+2\mathtt{q}\tilde{\Gamma}_{\rho\mathrm{r}}/\bar{\Gamma}_\mathrm{r}}{1+2\mathtt{q}} (k\tau_\mathrm{ini})^2 \mathcal{R}\, , \\
    \theta^\mathrm{ad}_\chi &= -\frac{23 + 30 \mathtt{q} \Gamma_{\rho\mathrm{r}} / \bar{\Gamma}_\mathrm{r} + 20 \mathtt{q} (1+2q_\chi) \Gamma_{\theta\mathrm{r}} / \bar{\Gamma}_\mathrm{r}}{180 (1+2\mathtt{q}) (3+4\mathtt{q})} k (k\tau_\mathrm{ini})^3 \mathcal{R} \, , \\
    \sigma^{\text{ad}}_\chi &= \frac{2}{45+90\mathtt{q}}(k\tau_\mathrm{ini})^2 \mathcal{R} \, ,
\end{align}
where 
\begin{equation}
    \mathtt{q} = \frac{a \bar{\mathcal{Q}}}{3 \mathcal{H} (\bar{\rho}_\chi+\bar{p}_\chi)}
\label{eq:fluid_quantities1}
\end{equation}
quantifies the efficiency of the energy transfer to DM in an expanding Universe, and
\begin{align}
    \bar{\Gamma}_\mathrm{r} &= \bar{\mathcal{Q}} /\bar{\rho}_\mathrm{r}
    \label{eq:fluid_quantities2} \\
    \Gamma_{\rho\mathrm{r}} &= \delta \mathcal{Q}_{0}/(\bar{\rho}_\mathrm{r} \delta_\mathrm{r}) \label{eq:fluid_quantities3} \\
    \Gamma_{\theta\mathrm{r}} &= \delta \mathcal{Q}_{1}  /((\bar{\rho}_\mathrm{r}+\bar{p}_\mathrm{r}) \theta_\mathrm{r}) \label{eq:fluid_quantities4}
\end{align}
are the dimensionless background energy exchange, perturbed energy exchange, and perturbed momentum exchange parameters, respectively.
For the purpose of finding an analytical solution of the system of equations, we assume that~$\mathtt{q}, \Gamma_{\rho\mathrm{r}}/\bar{\Gamma}_\mathrm{r}, \Gamma_{\theta\mathrm{r}}/\bar{\Gamma}_\mathrm{r}$ are constant during freeze-in.
The true time-dependence of these quantities is showed in section~\ref{sec:kodama-sasaki_formalism}, and is consistent with the assumption made here.
In the limit~$\mathtt{q}\to 0$, (relativistic) DM energy density redshifts as $a^{-4}$; in the limit $\mathtt{q} \to 1$, DM maintains a constant energy density.
The energy transfer is negligible in the $\mathtt{q} \to 0$ limit, and the initial conditions simplify to 
$\delta^{\text{ad}}_\chi = -(k\tau_\mathrm{ini})^2/3$, $\theta^{\text{ad}}_\chi = -23k^4\tau^3_\mathrm{ini}/540$, and~$\sigma^{\text{ad}}_\chi=2(k\tau_\mathrm{ini})^2/45$, consistent with the results expected from a subdominant, non-interacting, free-streaming species~\cite{bucher:initialconditions, kopp:generalizeddarkmatter}.

We identify the DM isocurvature mode as the mode that has constant overdensity~$\delta_\chi$ when the energy exchange is negligible (i.e., in the~$\mathtt{q}\to 0$ limit). 
In this case, at times when DM is relativistic, we have~$\lambda_\mathrm{iso}=0$, $\delta^{\text{iso}}_\chi=1$, $\theta^{\text{iso}}_\chi=k^2\tau_\mathrm{ini}/4$, and~$\sigma^{\text{iso}}_\chi=(k\tau_\mathrm{ini})^2/30$, matching the isocurvature initial conditions for a subdominant free-streaming species~\cite{bucher:initialconditions, kopp:generalizeddarkmatter}.
However, when the energy transfer is non-negligible, the isocurvature mode has a negative eigenvalue; hence, the mode is decaying as long as the energy exchange proceeds.
In this scenario, $\lambda_\mathrm{iso}=-4 \mathtt{q}$, which in the annihilation case with $\mathtt{q}^{\text{ann}} \approx 1/4$ simplifies to $\lambda_\mathrm{iso}\approx -1$ such that the DM isocurvature mode reads
\begin{align}
    h^\mathrm{iso} &= -\frac{\Gamma_{\rho\mathrm{r}}}{\bar{\Gamma}_\mathrm{r}} \varpi^2_\chi k\tau_\mathrm{ini} S_{\chi \mathrm{r}} \\
    \eta^\mathrm{iso} &= -\frac{3\Gamma_{\theta\mathrm{r}}+ \Gamma_{\rho\mathrm{r}}}{12\bar{\Gamma}_\mathrm{r}} \varpi^2_\chi k\tau_\mathrm{ini} S_{\chi \mathrm{r}} \\
    \delta^\mathrm{iso}_\mathrm{r} &= -\varpi_\chi S_{\chi \mathrm{r}} \\
    \theta^\mathrm{iso}_\mathrm{r} &= -(\varpi^2_\chi/4)S_{\chi \mathrm{r}} \\
    \delta^\mathrm{iso}_\chi &= (k\tau_\mathrm{ini})^{-1} S_{\chi,r} \\ 
    \theta^\mathrm{iso}_\chi &= (k/4) S_{\chi \mathrm{r}} \\
    \sigma^\mathrm{iso}_\chi &= \left(\frac{1}{30} - \frac{\Gamma_{\theta\mathrm{r}}+\Gamma_{\rho\mathrm{r}}}{10\bar{\Gamma}_\mathrm{r}} \varpi^2_\chi \right) k\tau_\mathrm{ini} S_{\chi \mathrm{r}} \, ,
\end{align}
where~$\varpi_\chi = H_0 \Omega_{\chi 0} / (\sqrt{\Omega_\mathrm{r0}}k)$, and~$H_0$, $\Omega_{\chi 0}$ and~$\Omega_\mathrm{r0}$ are the present-day Hubble expansion rate, DM relic abundance, and radiation relic abundance, respectively.
The amplitude of the isocurvature perturbation is~$S_{\chi \mathrm{r}}$.
In the decay case we have~$\mathtt{q}^\mathrm{d} \approx 3/4$ and
\begin{align}
    h^\mathrm{iso} &= -\frac{1}{5}\left(1+\frac{2}{3}\frac{\Gamma_{\rho\mathrm{r}}}{\bar{\Gamma}_\mathrm{r}} \right) \varpi_\chi (k\tau_\mathrm{ini})^3 S_{\chi \mathrm{r}} \\
    \eta^\mathrm{iso} &= - \frac{\varpi_\chi}{90} (k\tau_\mathrm{ini})^2 S_{\chi \mathrm{r}} \\
    \delta^\mathrm{iso}_\mathrm{r} &= -\varpi_\chi S_{\chi \mathrm{r}} \\
    \theta^\mathrm{iso}_\mathrm{r} &= -(\varpi^2_\chi/4)S_{\chi \mathrm{r}} \\
    \delta^\mathrm{iso}_\chi &= (k\tau_\mathrm{ini})^{-3} S_{\chi \mathrm{r}} \\ 
    \theta^\mathrm{iso}_\chi &= (k/4) (k\tau_\mathrm{ini})^{-2} S_{\chi \mathrm{r}} \\
    \sigma^\mathrm{iso}_\chi &= \frac{(k\tau_\mathrm{ini})^{-2}}{30} k\tau_\mathrm{ini} S_{\chi \mathrm{r}} \, ,
\end{align}
where~$\varpi_\chi = H_0 \Omega_{\chi 0} / (\sqrt{\Omega_\mathrm{r0}}k^3\tau^2_\mathrm{fi})$, where~$\tau_\mathrm{fi}$ is the conformal time at the end of the freeze-in.

\begin{figure}
    \centerline{
    \includegraphics[width=\columnwidth]{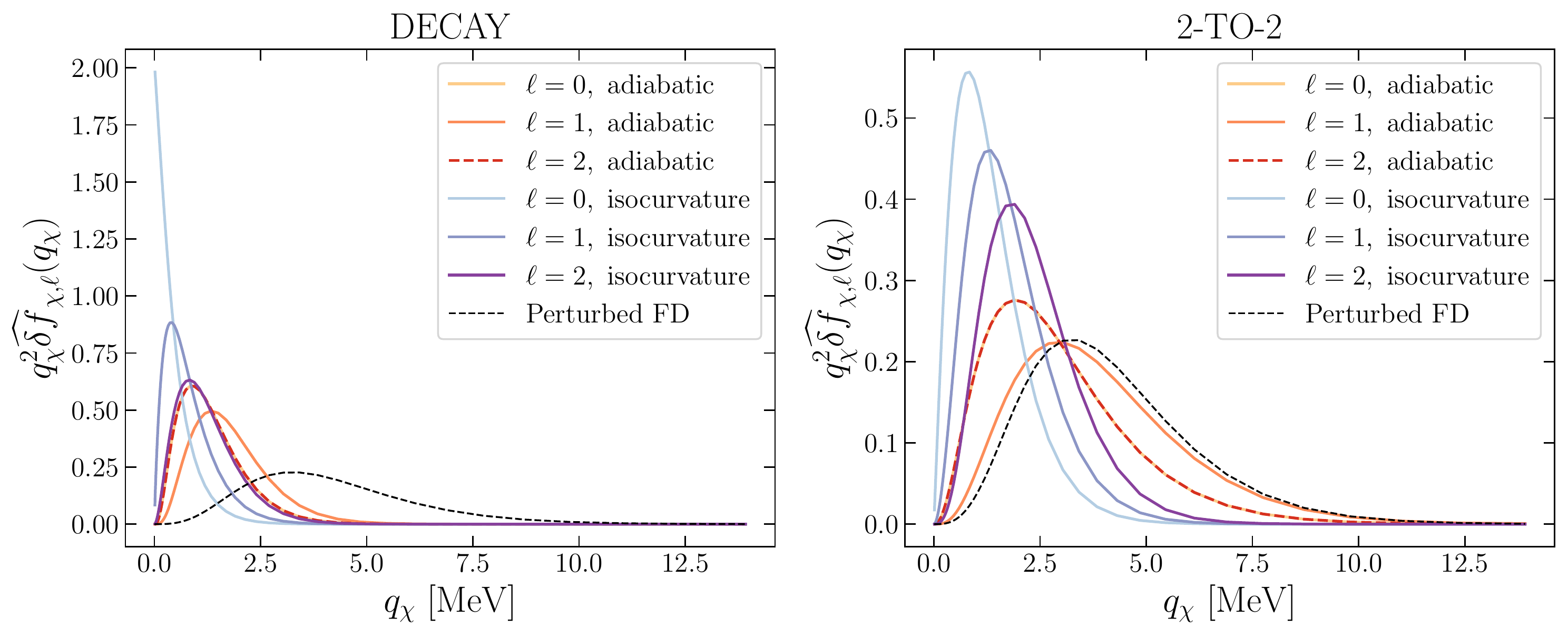}}
    \caption{Multipoles of the perturbed PSD of freeze-in DM for the decay (\textit{left panel}) and 2-to-2 annihilation (\textit{right panel}) scenarios as a function of comoving momentum. 
    We show the adiabatic and DM isocurvature modes for the decay case with $m_{Z'}=10\ \mathrm{GeV},\ m_\chi=4\ \mathrm{GeV}$ and the annihilation case with~$m_\chi=10\ \mathrm{MeV}$.
    The normalization for each PSD is set such that it integrates to 1, and the scale factor is normalized to be unity at~$T=1\ \mathrm{MeV}$.
    The dashed, black line represents the perturbed Fermi-Dirac (FD) distribution.}
    \label{fig:pt_psd}
\end{figure}

Given the initial conditions for the DM PSD and the evolution of the other perturbed variables, we can now solve the Boltzmann hierarchy in eq.~\eqref{eq:boltzmann_hierarchy_synchronous_gauge} for each individual mode. 
We show in figure~\ref{fig:pt_psd} the~$\ell=0,1,2$ multipoles of the adiabatic and DM isocurvature perturbed PSD for the decay and annihilation freeze-in cases at the end of the freeze-in process.
The qualitative difference between decay and annihilation case is the shape of~$\widehat{\delta f}_{\chi,0}$: the isocurvature mode of this PSD multipole does not evolve in time and thus remains imprinted in the features of the background PSD (cfr. eq.~\eqref{eq:perturbed_PSD_initial_considtions}) at~$\tau_\mathrm{ini}$.
In the decay case, at the initial time, the background PSD at small momenta~$m_\chi a(\tau_\mathrm{ini})\lesssim q_\chi \lesssim 1$ scales as~$\bar{f}_\chi\approx -q_\chi^{-2}\log q_\chi$, and thus~$q_\chi^2\widehat{\delta f}_{\chi,0} \propto -q_\chi^2 d\bar{f}_\chi/d\log q_\chi \approx -\log q_\chi$, explaining the growth in figure~\ref{fig:pt_psd}.
For momenta~$q_\chi \lesssim m_\chi a(\tau_\mathrm{ini})$, the perturbed PSD tend to zero.
On the other hand, in the annihilation case the growth of the background PSD at low momenta is less steep; hence, we recover a more familiar shape for $\widehat{\delta f}_{\chi,0}$.
As with the background PSD, the different multipoles (and hence the total perturbed PSD) are peaked at lower momenta with respect to the perturbed Fermi-Dirac distribution.


\section{Evolution of primordial perturbations}
\label{sec:kodama-sasaki_formalism}

\subsection{Kodama-Sasaki formalism}

We employ the formalism developed by Bardeen~\cite{bardeen:gaugeinvariantperturbations} and extended by Kodama and Sasaki (KS)~\cite{kodama:gaugeinvariantperturbations, hamazaki:gaugeinvariantperturbations} to study the evolution of the amplitudes for the adiabatic and isocurvature perturbations in our two-component universe, consisting of freeze-in DM and the SM bath.
In this framework the scalar metric perturbations are given in terms of four functions of conformal time and comoving coordinates ($A,B,H_L,H_T$), where the metric is  
\begin{equation}
    ds^2 = a^2(\tau)\left\{-(1+2A)d\tau^2 -2\frac{\partial B}{\partial x^i} dx^i d\tau + \left[ (1+2H_L)\delta_{ij} +2 \frac{\partial^2 H_T}{\partial x^i\partial x^j} \right] dx^i dx^j \right\} \, .
\end{equation}
The KS formalism consists of four differential equations quantifying the evolution of four gauge invariant variables: the total density perturbation $\Delta$, the total velocity $V$, the isocurvature perturbation amplitude $S_{\chi \mathrm{r}}$, and the relative velocity $V_{\chi \mathrm{r}}$. 
Our goal is to solve the KS equations throughout the DM freeze-in process in order to track the evolution of amplitude of the isocurvature mode $S_{\chi \mathrm{r}}$.
The gauge-invariant KS variables in our two fluid system are defined as
\begin{align}
    \bar{\rho}\Delta &=  \bar{\rho}_\chi \Delta_\chi + \bar{\rho}_\mathrm{r} \Delta_\mathrm{r} \\
    (\bar{\rho}+\bar{p})V &= (\bar{\rho}_{\chi} + \bar{p}_{\chi}) V_{\chi} + (\bar{\rho}_{\text{r}} + \bar{p}_{\text{r}}) V_{\text{r}} \\  
    \label{eq:isocurvature}
    S_{\chi \text{r}} &= \frac{\Delta_{\chi}}{1+w_{\chi}} - \frac{\Delta_{\text{r}}}{1+w_{\text{r}}} \\
    V_{\chi \text{r}} &= V_{\chi} - V_{\text{r}}\, ,
\end{align}
where
\begin{align}
\label{eq:KSvar1}
    \Delta_{\chi} &= \delta_\chi + 3 (1+w_\chi) (1-\mathtt{q}_\chi) \frac{\mathcal{H}}{k}(\theta/k-B) \\
    \label{eq:KSvar2}
    \Delta_{\text{r}}  &= \delta_{\text{r}} + 3 (1+w_{\text{r}})(1-\mathtt{q}_\mathrm{r})\frac{\mathcal{H}}{k}(\theta/k-B) \\
    \label{eq:KSvar3}
    V_\alpha &= \frac{\theta_\alpha}{k} - \frac{1}{k}\frac{dH_T}{d\tau} \\ (\bar{\rho}+\bar{p})\theta &= (\bar{\rho}_{\chi}+\bar{p}_{\chi}) \theta_{\chi} + (\bar{\rho}_{\text{r}}+\bar{p}_{\text{r}}) \theta_{\text{r}} \, .
\label{eq:KS_gaugeinvariant_elementary_variables}
\end{align}
The curvature perturbation~$\mathcal{R}$ is related to the total density perturbation~$\Delta = -(2/3)(k\tau)^2\mathcal{R}$.
During freeze-in the Universe is radiation dominated such that $q_\mathrm{r} \ll q_{\chi}$; hence, we neglect $q_\mathrm{r}$ when solving the KS equations.
Furthermore, the radiation sector has equation of state~$w_\mathrm{r} = 1/3$ and adiabatic sound speed~$c^2_{a,\mathrm{r}} = 1/3$; therefore, since we have~$\bar{\rho}_{\chi} \ll \bar{\rho}_\mathrm{r}$, the total equation of state and adiabatic sound speed of the multi-fluid system is also~$w \approx c_a^2 \approx 1/3$.
Performing a change of variables from $\tau$ to $x \equiv k \tau$, the KS system of four differential equations simplifies to 
\label{eq:KSfull}
\begin{align}
    \label{eq:KSfull1}
    \frac{d\Delta}{dx} = & \frac{\Delta - \frac{2}{3}\Pi}{x} -\frac{4}{3} V \\
    \label{eq:KSfull2}
    \frac{dV}{dx} = & - \frac{V}{x} - \left[\frac{3}{2x^2} - \frac{1}{4}\right]\Delta + \frac{1}{4}(\Gamma_\mathrm{int}+\Gamma_\mathrm{rel}) - \left[\frac{1}{x^2} + \frac{1}{6}\right]\Pi \\
    \label{eq:KSfull3}
    \frac{dS_{\chi\text{r}}}{dx} =& - \frac{3}{x}(1+c^2_{a,{\chi}})\mathtt{q}_{\chi} S_{\chi \text{r}}  - V_{\chi \text{r}} - \frac{3}{x} \frac{w_\chi}{1+w_\chi}\Gamma_\chi  + \frac{3}{x}E_{\chi \text{r}} \\
    & - \frac{3}{4x}\left[\mathtt{q}_\chi + 3(1+c^2_{a,\chi})\mathtt{q}_\chi \right]\Delta - \frac{3\mathtt{q}_{\chi}}4{x}\left[\Gamma_\mathrm{int}+\Gamma_\mathrm{rel} - \frac{2}{3}\Pi \right] \nonumber \\
    \label{eq:KSfull4}
    \frac{dV_{\chi \text{r}}}{dx} =& - \frac{3}{x}\left[\frac{1}{3}-c^2_{a,\chi} +  (1+c^2_{a,\chi})\mathtt{q}_\chi \right] V_{\chi \text{r}} 
    + \left(c^2_{a,\chi} - \frac{1}{3}\right)\frac{3\Delta}{4} + \frac{1}{x}F_{\chi \text{r}}  \\ & + \left[\frac{w_\chi}{1+w_\chi}\Gamma_\chi - 
    \frac{2}{3} \frac{w_\chi}{1+w_\chi}\Pi_\chi  \right] + c^2_{a,\chi} S_{\chi \text{r}} \, ,
\end{align}
where we define the dimensionless perturbed energy and velocity exchange parameters in the KS formalism in terms of the integrated background and perturbed collision terms we derive in sections~\ref{subsec:background_cosmology} such that 
\begin{align} 
\label{eq:energy}
    E_{\chi \text{r}} = \mathcal{E}_\chi - \mathcal{E}_\mathrm{r} \approx \mathcal{E}_{\chi} =& \mathtt{q}_{\chi}\left(\frac{\delta \mathcal{Q}_{0}} {\bar{\mathcal{Q}}} - \frac{1}{\mathcal{H}\bar{\mathcal{Q}}}\frac{d\bar{\mathcal{Q}}}{d\tau}\frac{\mathcal{H}(\theta/k-B)}{k} \right)\, , \\
    F_{\chi \text{r}} = F_{\chi} -F_\mathrm{r} \approx   F_\chi =& \frac{\delta \mathcal{Q}_{1}-\bar{\mathcal{Q}}\theta}{a\mathcal{H}k(\bar{\rho}_\chi+\bar{p}_\chi)}\, .
\end{align} 
Additionally, the intrinsic non-adiabatic pressure perturbation $\Gamma_{\text{int}}$, the relative entropy perturbation $\Gamma_{\text{rel}}$, and the anisotropic stress $\Pi$ are
\begin{align}
    \bar{p}\Gamma = & \bar{p}\Gamma_\mathrm{int} + \bar{p}\Gamma_\mathrm{rel}\, , \label{eq:KS_fluid_properties1} \\
    \bar{p}\Pi = & \bar{p}_\chi \Pi_\chi  + \bar{p}_\text{r} \Pi_\text{r}\approx \frac{3}{2}   (\bar{\rho}_\chi+\bar{p}_\chi)\sigma_\chi
    \, , \\
    \bar{p}\Gamma_\mathrm{int} = & \bar{p}_\chi \Gamma_\chi + \bar{p}_\text{r} \Gamma_\text{r} \approx  \left(\delta p_{\chi} - c^2_{a,\chi}\delta \rho_\chi\right) \, , \\
    \bar{p}\Gamma_\mathrm{rel} =  & \left(c^2_{a,\chi}-c^2_{a,\mathrm{r}}\right)(\bar{\rho}_\chi+\bar{p}_\chi) \left[S_{\chi \mathrm{r}} + \mathtt{q}_\chi \frac{\Delta}{1+w} \right]\, ,
    \label{eq:KS_fluid_quantities4}
\end{align}
respectively.
The anisotropic stress $\Pi_\mathrm{r}$ and internal pressure perturbation $\Gamma_\mathrm{r}$ of the thermal radiation are negligible due to the fast interactions in the SM bath, which maintain equilibrium and suppress deviations from a perfect fluid.
Therefore, calculating the DM fluid macroscopic properties is sufficient to quantify the global properties in equations~\eqref{eq:KS_fluid_properties1}-\eqref{eq:KS_fluid_quantities4}.  

\begin{figure}[t]
    \centerline{
    \includegraphics[width=\columnwidth]{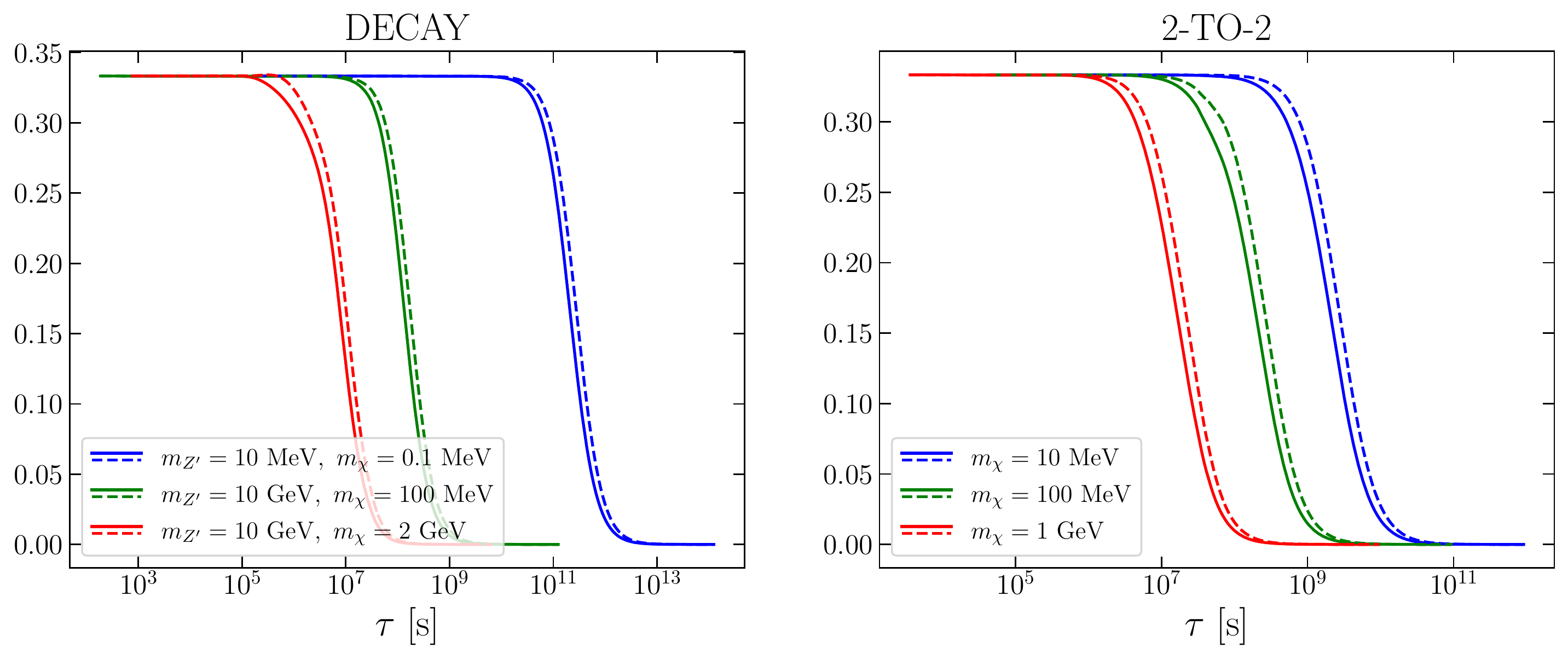}}
    \caption{Evolution of the equation of state~$w_\chi$ (\textit{solid lines}) and adiabatic sound speed~$c^2_{a,\chi}$ (\textit{dashed lines}) for the decay (\textit{left panel}) and 2-to-2 annihilation (\textit{right panel}) freeze-in scenarios.}
    \label{fig:bg_fluid_properties}
\end{figure}

Using the solutions of the perturbed and background evolution equations for the DM PSD, we calculate the evolution of the quantities in eqs.~\eqref{eq:KS_fluid_properties1}-\eqref{eq:KS_fluid_quantities4}, the DM equation of state~$w_{\chi}$, and the adiabatic sound speed~$c^2_{a,\chi}$.
In figure~\ref{fig:bg_fluid_properties}, we show the evolution of $w_{\chi}$ and~$c^2_{a,\chi}$ as a function of conformal time~$\tau$ for various DM masses for freeze-in through decay and annihilation. 
As expected, the DM is relativistic at early times with an equation of state of $w_{\chi} =1/3$, and it eventually transitions to cold DM with $w_{\chi}(\tau_{\text{fin}}) = 0$.
The adiabatic sound speed tracks the evolution of the equation of state but with a small time delay.  

\begin{figure}[t]
    \centerline{
    \includegraphics[width=\columnwidth]{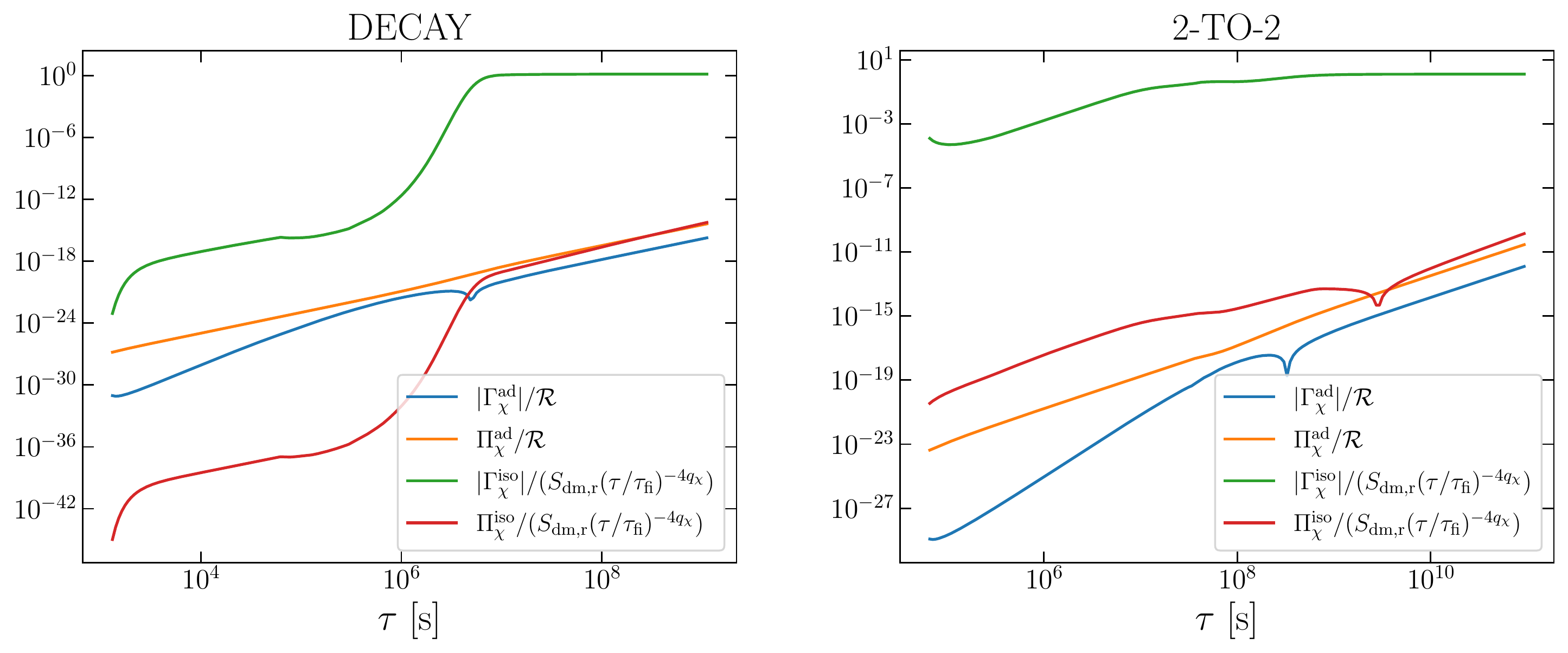}}
    \caption{Evolution of the DM non-adiabatic pressure~$\Gamma_\chi$ and anisotropic stress~$\Pi_\chi$ perturbations for the DM adiabatic and isocurvature modes for~$k=10^{-2}\ \mathrm{Mpc}^{-1}$.
    Curves are normalized by the total amplitude of their respective mode to represent the intrinsic value.
    We set $m_{Z'}=10\ \mathrm{GeV}$ and $m_\chi=4\ \mathrm{GeV}$ for the decay case (\textit{left panel}) and $m_\chi=1\ \mathrm{GeV}$ for the 2-to-2 annihilation case (\textit{right panel}).}
    \label{fig:pt_fluid_properties}
\end{figure}

In figure~\ref{fig:pt_fluid_properties} we show the evolution of DM internal pressure and anisotropic stress fluid properties for both the adiabatic and the isocurvature mode.
We observe that all these fluid properties are suppressed during the DM formation epoch by powers of~$k\tau$.
In particular, we find that during DM formation, the non-adiabatic pressure perturbations evolve as~$|\Gamma^\mathrm{ad}_\chi|/\mathcal{R} \propto (k\tau)^4$ and~$|\Gamma^\mathrm{iso}_\chi|/(S_{\chi \mathrm{r}}(\tau/\tau_\mathrm{fi})^{-4\mathtt{q}_\chi})\propto (k\tau)^2$, which are compatible with the theoretical scaling with time that can be estimated as
\begin{equation}
    \Gamma^\mathrm{ad,iso}_\chi = \int \frac{2d^3q}{(2\pi)^3} \left(\frac{q^2}{3\epsilon}-c^2_{a,\chi}\epsilon\right) \widehat{\delta f}^\mathrm{ad,iso}_{\chi,0} \approx -\int d^3q \frac{q}{3} \frac{m^2_\chi a^2}{q^2} \widehat{\delta f}^\mathrm{ad,iso}_{\chi,0},
\end{equation}
where we keep only the leading order in expanding for~$q_\chi\gg m_\chi a$.
Therefore the expected scaling with time of~$\Gamma$ is given by the scaling of~$\delta_\chi$ times the additional factor~$a^2\propto \tau^2$.
On the other hand, during freeze-in, when~$\Pi_\chi \approx \sigma_\chi$, the anisotropic stress perturbation evolves as~$\Pi^\mathrm{ad}_\chi/\mathcal{R} \propto (k\tau)^2$ and~$|\Pi^\mathrm{iso}_\chi|/(S_{\chi \mathrm{r}}(\tau/\tau_\mathrm{fi})^{-4\mathtt{q}_\chi})\propto (k\tau)^2$ as suggested by the mode evolution described in section~\ref{subsec:initial_conditions_perturbations}.
These behaviours are common to both models.
The relative entropy perturbation~$\Gamma_\mathrm{rel}$ vanishes during freeze-in, since~$c^2_{a,\chi}=c^2_{a,\mathrm{r}}$. 
Note that all these properties are gauge-invariant by construction; hence, they can be computed in any gauge of choice when solving the perturbed PSD Boltzmann hierarchy.

\begin{figure}[t]
    \centerline{
    \includegraphics[width=\columnwidth]{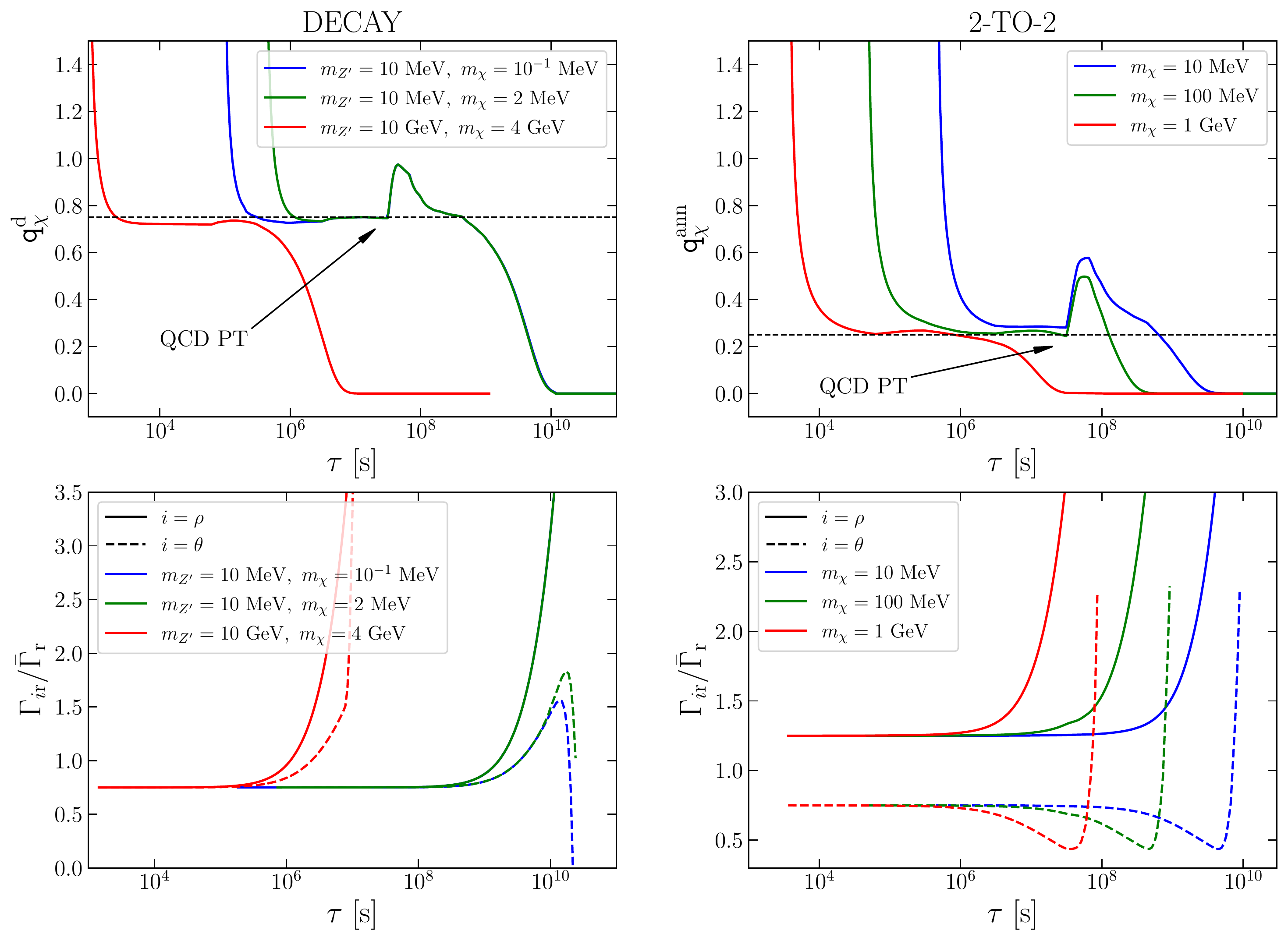}}
    \caption{\textit{Top panels}: Relative background energy exchange parameter~$\mathtt{q}$ for the decay (\textit{left panel}) and 2-to-2 annhilation (\textit{right panel}) freeze-in models.
    Dashed lines corresponds to the values~$\mathtt{q}^\mathrm{d}=3/4$ and~$\mathtt{q}^\mathrm{ann}=1/4$ which we used for analytical estimate of sections~\ref{subsec:initial_conditions_perturbations} and~\ref{subsec:initial_isocurvature_evolution}.
    The kinks correspond to the beginning of the QCD phase transition (PT).
    \textit{Bottom panels}: Perturbed energy~$\Gamma_{\rho\mathrm{r}}/\bar{\Gamma}_\mathrm{r}$ (\textit{solid lines}) and momentum~$\Gamma_{\theta\mathrm{r}}/\bar{\Gamma}_\mathrm{r}$ (\textit{dashed lines}) exchange coefficients.}
\label{fig:energy_exchange_properties}
\end{figure}

We turn our attention to the non-zero energy exchange term $E_{\chi \textrm{r}}$. 
For the freeze-in models we consider, the energy transfer term is a function of temperature of the thermal bath such that $\mathcal{Q}=\mathcal{Q}_{\chi}(\bar{T})$, where $\bar{T}$ is determined by the radiation energy density $\bar{\rho}_\mathrm{r} = \frac{30}{\pi^2} g_{*}(\bar{T}) \bar{T}^4$.
The relative energy exchange term appearing in eq.~\eqref{eq:energy} can be rewritten as
\begin{equation} 
\label{eq:pertenergy}
    \frac{\delta \mathcal{Q}_{0}}{\bar{\mathcal{Q}}} = \frac{d\log \mathcal{Q}_\chi}{d\log \bar{\rho}_\mathrm{r}} \delta_\mathrm{r} = \frac{d\log \mathcal{Q}_\chi}{\mathcal{H}d\tau}  \left(\frac{d\log \bar{\rho}_\mathrm{r}}{\mathcal{H}d\tau}\right)^{-1} \left[\Delta_{\mathrm{r}} + \frac{d\log \bar{\rho}_\mathrm{r}}{\mathcal{H}d\tau} \frac{\mathcal{H}(\theta/k-B)}{k} \right] \, .
\end{equation}
Therefore, eq.~\eqref{eq:energy} simplifies to 
\begin{equation}
\label{eq:Eexchange}
    E_{\chi \textrm{r}} = (1+w_\mathrm{r}) \mathtt{q}_\chi \frac{d\log \mathcal{Q}_\chi}{d\log \bar{\rho}_\mathrm{r}} \left[ \frac{\Delta}{1+w} -  \frac{\bar{\rho}_\chi+\bar{p}_\chi}{\bar{\rho}+\bar{p}} S_{\chi \textrm{r}}\right]\, .
\end{equation}
In the case zero initial DM isocurvature $S_{\chi \textrm{r}}=0$, consistent with single field inflation, the energy exchange term vanishes in the large scale limit $x \to 0$ (since $\Delta \sim x^2$) and thus does not source isocurvature.

In figure~\ref{fig:energy_exchange_properties} we show the evolution of $\mathtt{q}^\mathrm{d}_\chi$, $\mathtt{q}^{\text{ann}}_\chi$, and $\Gamma_{\rho\mathrm{r}}/\bar{\Gamma}_\mathrm{r},\Gamma_{\theta\mathrm{r}}/\bar{\Gamma}_\mathrm{r}$ defined in eqs. \eqref{eq:fluid_quantities1}-\eqref{eq:fluid_quantities4} for a selection of DM masses in our two benchmark models.
Regarding the parameters~$\mathtt{q}$ for decay and annihilation, we observe that their values do not vary significantly during freeze-in, except during the QCD phase transition at~$\tau\simeq 10^8\ \mathrm{s}$ for reasons discussed in section~\ref{subsec:background_cosmology}.
The initial steep phase of the curve corresponds to DM rapidly approaching its background attractor solution, since we assume that at initial time~$\bar{f}_\chi(\tau_\mathrm{ini})=0$.
Conversely, if a sizeable fraction of DM is already present at initial time, we would have observed~$\mathtt{q}_\chi\to 0$ also at early times.
On the other hand, we observe that our treatment of the~$\Gamma_{\rho\mathrm{r}}/\bar{\Gamma}_\mathrm{r},\Gamma_{\theta\mathrm{r}}/\bar{\Gamma}_\mathrm{r}$ coefficients in section~\ref{subsec:initial_conditions_perturbations} as effective constants is well-justified in the regime where~$\bar{T} \gg m_{Z'}$ and~$\bar{T}\gg m_\chi,m_e$. 
When the temperature drops closer to the mass scales of the particles involved in freeze-in, the evolution of these quantities is affected by the choice of mass parameters.
However, note that these differences appear only during the final stages of freeze-in, when energy and momentum transfer start becoming inefficient.


\subsection{Dark matter freeze-in with initial isocurvature}
\label{subsec:initial_isocurvature_evolution}

Generically, in a single-field slow-roll model of inflation, only the adiabatic mode is excited.
The perturbation in the inflaton field~$\phi$ can be expressed in terms of a time shift~$\delta t$ of the background scalar field such that~\cite{lyth:inflationreview}
\begin{equation}
    \delta \phi = \frac{d\phi}{dt} \delta t \, .
\end{equation}
After inflation, perturbations in the spatial distribution of different species are connected by 
\begin{equation}
    \delta t = \frac{\delta \rho_{\chi}}{\displaystyle \frac{d\bar{\rho}_\chi}{d\tau}} = \frac{\delta \rho_{\text{r}}}{\displaystyle \frac{d\bar{\rho}_\text{r}}{d\tau}} \, .
\label{eq:single_clock}
\end{equation}
which implies $S_{\chi \text{r}} \approx 0$ due to the energy density continuity equation.
Any contributions to $S_{\chi \text{r}}$ from the coupled evolution with $\mathcal{R}$ that vanish in the $k\tau \to 0$ limit are not fundamental excitations of the isocurvature mode and thus are not observable~\cite{kodama:gaugeinvariantperturbations}. 

However, in the presence of multiple fields during the inflationary era, the DM isocurvature mode can be excited. 
For example, if a second subdominant scalar field is present during inflation, its fluctuations have non-zero power spectrum (e.g., see refs.~\cite{LINDE1985375, Polarski:1994rz, linde:isocurvaturefrominflation, Garcia-Bellido:1995hsq, 
axenides:axionisocurvature, seckel:axionisocurvature, turner:axionisocurvature}).
We consider a scenario in which a fraction of DM is present at the end inflation with an excited isocurvature mode with amplitude $S^{\text{ini}}_{\chi \text{r}} > 0$.
The remaining fraction of DM is produced post inflation via freeze-in.
Note that we still assume there is a single species of DM particle.

To analyze this scenario, we solve the KS equations with the initial conditions $S^{\text{ini}}_{\chi \text{r}} = \mathcal{R}$ and initial fraction of DM abundance $\mathcal{F}_\mathrm{ini}$ (see section~\ref{subsec:background_cosmology}). 
In the context of reheating, efficient energy exchange has been shown to lead to an exponential decay of isocurvature~\cite{Weinberg:2004kf}. 
This scenario would apply, for example, if DM were a WIMP in thermal equilibrium with the SM bath in the early Universe. 
Any initial DM isocurvature would be erased due to DM becoming part of the thermal bath. 

\begin{figure}[t]
    \centerline{
    \includegraphics[width=\columnwidth]{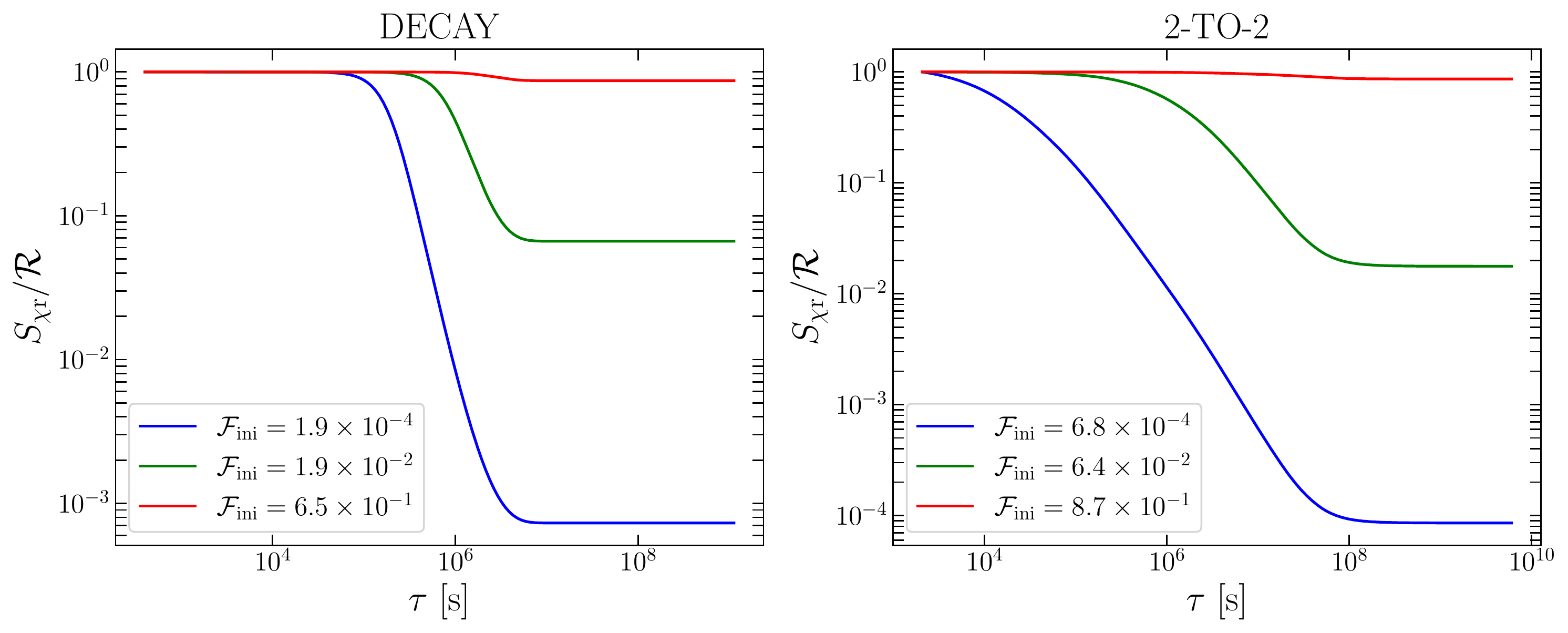}}
    \caption{Isocurvature amplitude evolution, normalized by the curvature amplitude evolution, for the decay (\textit{left panel}) and 2-to-2 annihilation (\textit{right panel} freeze-in scenarios.
    We set $m_{Z'}=10\ \mathrm{GeV}$ and $m_\chi=2\ \mathrm{GeV}$ for the decay case and $m_\chi=1.6\ \mathrm{GeV}$ for the 2-to-2 annihilation case.
    We show the evolution for different comoving DM abundances~$\mathcal{F}_\mathrm{ini}$.
    Solid and dashed lines correspond to~$S^\mathrm{ini}_{\chi \mathrm{r}} = \mathcal{R}$ and~$S^\mathrm{ini}_{\chi \mathrm{r}} = 10^{-1} \mathcal{R}$, respectively.}
    \label{fig:isocurvature_evolution}
\end{figure}

However, in the freeze-in scenario, the interactions are not efficient enough to erase the initial isocurvature, as shown in figure~\ref{fig:isocurvature_evolution}.
We find that the isocurvature perturbation, despite being diluted, could still have an impact on late time observables.
The evolution of the energy and momentum source terms~$E_{\chi \mathrm{r}}$ and~$F_{\chi \mathrm{r}}$ depends not only on the properties of DM, but also on the relative size of the primordial curvature and isocurvature perturbations.
For our benchmark models we have that
\begin{equation}
    \begin{aligned}
    E_{\chi \mathrm{r}} &\simeq -\frac{4}{3} \mathtt{q}_\chi \frac{\Gamma_{\rho\mathrm{r}}}{\bar{\Gamma}_\mathrm{r}} \left[\frac{3}{4}\Delta -  \frac{\bar{\rho}_\chi}{\bar{\rho}_\mathrm{r}} S_{\chi \mathrm{r}} \right] \propto
    \left\lbrace \begin{matrix}
    x^2\mathcal{R} \qquad \mathrm{decay} \\
    xS_{\chi \mathrm{r}} \qquad \mathrm{ann.}\\
    \end{matrix} \right.
    \, ,\\
    F_{\chi \mathrm{r}} &\simeq 3\mathtt{q}_\chi\left[\left(\frac{4\Gamma_{\theta\mathrm{r}}}{3\bar{\Gamma}} - 1\right) V - \frac{4\Gamma_{\theta\mathrm{r}}}{3\bar{\Gamma}} \frac{\bar{\rho}_\chi+\bar{p}_\chi}{\bar{\rho}+\bar{p}} V_{\chi \mathrm{r}} \right] \propto x\mathcal{R} \qquad \mathrm{decay,\ ann.},
    \end{aligned}
\end{equation}
at leading order during radiation domination.
Both source terms are suppressed on super-horizon scales.
Since~$\bar{\rho}^\mathrm{d}_\chi/\bar{\rho}_\mathrm{r} \propto x^3$ and~$\bar{\rho}^\mathrm{ann}_\chi/\bar{\rho}_\mathrm{r} \propto x$, the term proportional to~$\Delta$ dominates in the decay case, while the isocurvature term dominates in the annihilation scenario, assuming that curvature and isocurvature perturbations have comparable amplitudes.
At leading order the evolution of isocurvature is governed by 
\begin{equation}
    \frac{dS_{\chi \mathrm{r}}}{dx} \approx - \frac{3}{x} (1+c^2_{a,\chi}) \mathtt{q}_{\chi} S_{\chi \mathrm{r}} \, .
\label{eq:freezein_isocurvature_evolution}
\end{equation}
Thus, we have~$S_{\chi \mathrm{r}}\propto \tau^{-3}$ for the decay case with $\mathtt{q}^{\text{d}}_{\chi} \approx 0.75$ and~$S_{\chi \mathrm{r}}\propto \tau^{-1}$ for the annihilation case with $\mathtt{q}^{\text{ann}}_{\chi} \approx 0.25$.
A rescaling of the initial value of the isocurvature perturbation~$S^\mathrm{ini}_{\chi \mathrm{r}}$ simply corresponds to vertical shift of the curves.
In conclusion, the resulting isocurvature, which is seeded during inflation, does not vanish on large scales and is subject to constraints from the CMB.
The degree of statistical correlation between the isocurvature and adiabatic fluctuation is not affected by freeze-in, since in this scenario, the two perturbations do not mix.


\section{Current and future constraints on isocurvature}
\label{sec:detectability_forecast}


Over the past two decades, CMB experiments have favored cosmological initial conditions that are adiabatic and very nearly Gaussian~\cite{peiris:wmapinflation, komatsu:wmapinflation, ade:planckinflationI, ade:plankinflationII, akrami:planckinflation}. 
The \textit{Planck} collaboration constrains the primordial isocurvature fraction parameter $\beta_\mathrm{iso}$ and the degree of correlation $\delta$ between the curvature and isocurvature perturbations, which are defined as~\cite{ade:planckinflationI, ade:plankinflationII, akrami:planckinflation}
\begin{equation}
    \beta_\mathrm{iso}(k) = \frac{\mathcal{P}_{SS}(k)}{\mathcal{P}_{\mathcal{R}\mathcal{R}}(k)+\mathcal{P}_{SS}(k)}, \qquad \cos\delta = \frac{\mathcal{P}_{\mathcal{R}S}}{\sqrt{\mathcal{P}_{\mathcal{R}\mathcal{R}}+\mathcal{P}_{SS}}} \, .
\end{equation} 
We forecast the reach of future CMB measurements with Simons Observatory (SO)~\cite{ade:SOwhitepaper} and CMB-S4~\cite{abazajian:cmbs4whitepaper} for three classes of a  1-parameter extension of the~$\Lambda$CDM model: totally correlated~($\cos\delta=1$), anti-correlated~($\cos\delta=-1$), and uncorrelated~($\cos\delta=0$) DM isocurvature perturbations.
We fix the spectral tilt of isocurvature power spectrum to be~$n_{II}=1$ (uncorrelated scenario) or~$n_{II}=n_{\mathcal{R}\mathcal{R}}$ (totally correlated/anti-correlated scenarios), making the~$\beta_\mathrm{iso}$ parameter fundamentally scale-independent.
If the spectral tilt of the isocurvature power spectrum is left free to vary, data prefer~$n_{\mathcal{I}\mathcal{I}}\gtrsim 1$, which corresponds to a weaker bound on~$\beta_\mathrm{iso}$~\cite{ade:planckinflationI, ade:plankinflationII, akrami:planckinflation}.

Using the the Boltzmann code \texttt{CLASS}~\cite{blas:class}, we compute the minimum level of isocurvature perturbations detectable by next-generation experiments by through their impact on the temperature and polarization anisotropies of the CMB.
We perform a Fisher matrix analysis to obtain estimates of the constraining power. 
Both SO and CMB-S4 are located in the Southern Hemisphere, with fractional sky coverage of~$f_\mathrm{sky}=0.4$.
For SO we consider the goal configuration, characterized by a white instrumental noise for temperature of~$\mathcal{N}^{TT}=6.3\ \mu K \mathrm{-arcmin}$.
For CMB-S4 we assume~$\mathcal{N}^{TT}=1\ \mu K\mathrm{-arcmin}$.
The expected spatial resolution for SO is~$\theta_\mathrm{FWHM}=1.4\ \mathrm{arcmin}$, and we use the conservative value of~$\theta_\mathrm{FWHM}=3\ \mathrm{arcmin}$ for CMB-S4.
For both experiments, the polarization noise is taken to be~$\mathcal{N}^{EE}=\sqrt{2}\mathcal{N}^{TT}$.
The range of multipoles of interest is~$\ell=[30,3000]$ for SO and~$\ell_{T}=[30,3000]$ and $\ell_{E}=[30,5000]$ for temperature and polarization for CMB-S4, respectively. 

\begin{table}[ht]
    \centerline{
    \begin{tabular}{|c|c|c|c|c|c|c|}
    \hline
     & CMB probe & \textit{Planck} (TEL) & SO (TE) & SO (TEL) & CMB-S4 (TE) & CMB-S4 (TEL) \\
    \hline
    \hline
    \multirow{3}{*}{$\cos\delta=1$} & $\beta_\mathrm{iso}$ & $9.5\times 10^{-4}$ & $1.5 \times 10^{-4}$ & $1.4\times 10^{-4}$ & $1.4\times 10^{-4}$ & $6.7\times 10^{-5}$ \\
     & $\mathcal{F}^\mathrm{d}_\mathrm{ini}$ & $9\times 10^{-3}$ & $3\times 10^{-3}$ & $3\times 10^{-3}$ & $3\times 10^{-3}$ & $2\times 10^{-3}$ \\
     & $\mathcal{F}^\mathrm{ann}_\mathrm{ini}$ & $4\times 10^{-2}$ & $2\times 10^{-2}$ & $2\times 10^{-2}$ & $2\times 10^{-2}$ & $1.5\times 10^{-2}$ \\
    \hline
    \hline
    \multirow{3}{*}{$\cos\delta=0$} & $\beta_\mathrm{iso}$ & $3.8\times 10^{-2}$ & $1.7 \times 10^{-2}$ & $1.7\times 10^{-2}$ & $1.5\times 10^{-2}$ & $8.8\times 10^{-3}$ \\
     & $\mathcal{F}^\mathrm{d}_\mathrm{ini}$ & $0.12$ & $0.07$ & $0.07$ & $0.06$ & $0.04$ \\
     & $\mathcal{F}^\mathrm{ann}_\mathrm{ini}$ & $0.22$ & $0.15$ & $0.15$ & $0.14$ & $0.11$ \\
    \hline
    \hline
    \multirow{3}{*}{$\cos\delta=-1$} & $\beta_\mathrm{iso}$ & $10.7\times 10^{-4}$ & $1.5 \times 10^{-4}$ & $1.4\times 10^{-5}$ & $1.4\times 10^{-4}$ & $7.1\times 10^{-5}$ \\
     & $\mathcal{F}^\mathrm{d}_\mathrm{ini}$ & $9\times 10^{-3}$ & $3\times 10^{-3}$ & $3 \times 10^{-3}$ & $3\times 10^{-3}$ & $2\times 10^{-3}$ \\
     & $\mathcal{F}^\mathrm{ann}_\mathrm{ini}$ & $4\times 10^{-2}$ & $2\times 10^{-2}$ & $2\times 10^{-2}$ & $2\times 10^{-2}$ & $1.5\times 10^{-2}$ \\
    \hline
    \end{tabular}}
    \caption{Current upper limits and forecasted sensitivity on~$\beta_\mathrm{iso}$ for totally correlated~($\cos\delta=1$), uncorrelated~($\cos\delta=0$), and totally anti-correlated~($\cos\delta=-1$) DM isocurvature perturbations.
    We consider temperature and polarization anisotropies alone (TE), as well as including lensing (TEL).
    Values of~$\beta_\mathrm{iso}$ are quoted at the~$95\%\ \mathrm{CL}$ both for \textit{Planck}~\cite{akrami:planckinflation} and for the forecasted values upper bounds. 
    We also include derived upper bounds and forecasted sensitivity on the initial DM abundance for decay~$\mathcal{F}^\mathrm{d}_\mathrm{ini}$ and annihilation~$\mathcal{F}^\mathrm{ann}_\mathrm{ini}$ DM freeze-in.
    In the decay case, we consider~$m_{Z'}=10\ \mathrm{GeV}$ and~$m_\chi=2\ \mathrm{GeV}$.
    In the 2-to-2 annihilation case, we set~$m_\chi =100 \ \mathrm{MeV}$.}
    \label{tab:isocurvature_constraints}
\end{table}

Existing constraints for~$\beta_\mathrm{iso}$ and the results of our forecast are reported in table~\ref{tab:isocurvature_constraints}.
Improvements over \textit{Planck} for upcoming experiments are driven mainly by improvements in the polarization noise and provide up to one order of magnitude tighter upper bounds on the amplitude of the isocurvature power spectrum.
Given the time evolution of the isocurvature perturbation in eq.~\eqref{eq:freezein_isocurvature_evolution}, we can recast the final value of the isocurvature perturbation in terms of the initial DM abundance as
\begin{equation}
    S^\mathrm{fin}_{\chi \mathrm{r}} = \mathcal{A} \mathcal{F}_\mathrm{ini} \left(\frac{S^\mathrm{ini}_{\chi \mathrm{r}}}{\mathcal{R}}\right) \mathcal{R}, 
\end{equation}
where~$\mathcal{A}$ is numerical coefficient determined by the full evolution of the KS equations.
Therefore, constraints on~$\beta_\mathrm{iso}$ can also be recast in terms of constraints of initial DM abundance as
\begin{equation}
    \mathcal{F}_\mathrm{ini} \lesssim \frac{1}{\mathcal{A}(S^\mathrm{ini}_{\chi \mathrm{r}}/\mathcal{R})} \frac{\beta_\mathrm{iso}}{1-\beta_\mathrm{iso}}.
\label{eq:fini_upper_bound}
\end{equation}
We also report upper bounds on the initial DM abundance in table~\ref{tab:isocurvature_constraints} for the~$m_{Z'}=10\ \mathrm{GeV},\ m_\chi=2\ \mathrm{GeV}$ decay case and the~$m_\chi =100 \ \mathrm{MeV}$ annihilation scenario.
For our choice of~$S^\mathrm{ini}_{\chi \mathrm{r}}/\mathcal{R}=1$,  we find upper bounds on the primordial DM abundance that are~$\mathcal{F}_\mathrm{ini}\lesssim \mathcal{O}(10^{-2})$ at~$95\%$ CL for totally correlated/anti-correlated isocurvature perturbations and~$\mathcal{F}_\mathrm{ini}\lesssim \mathcal{O}(10^{-1})$ at~$95\%$ CL for uncorrelated ones, when considering temperature, polarization, and lensing anisotropies. 
Upper bounds for the decay case are a factor~$\mathcal{O}(2-5)$ tighter than the annihilation scenario, and they are only slightly sensitive to the exact values of~$m_{Z'}$ and~$m_\chi$.
We find that the upper bounds from SO and CMB-S4 are very similar and are typically a factor few better than those derived from \textit{Planck}.
Moreover, upper bounds on the primordial DM abundance for other points of the parameter space differ by less than a factor 2 from the reported values.
If the initial isocurvature-to-curvature ratio is different from 1, upper bounds on $f_\textrm{ini}$ scale according to eq.~\eqref{eq:fini_upper_bound}; for instance, if~$S^\mathrm{ini}_{\chi \mathrm{r}}/\mathcal{R}=10^{-1}$, upper bounds are ten times weaker, regardless of the degree of correlation between isocurvature and curvature perturbations.

In this work we assumed for simplicity that primordial isocurvature has the same scale dependence of the curvature perturbation.
If they possess different scale dependencies, the constraints are model-dependent; however, we speculate that our results remain valid for $n_\mathcal{R}\sim n_{S_{\chi \mathrm{r}}}$ assuming that curvature and isocurvature power spectra have the same pivot scale, since the constraints mainly come from low multipoles, due to the high-$\ell$ suppression of temperature fluctuations in the isocurvature mode~\cite{hu:analyticcmbanisotropiesI, hu:analyticcmbanisotropiesII}.
On the other hand, as suggested by the \textit{Planck} analysis~\cite{ade:planckinflationI, ade:plankinflationII, akrami:planckinflation}, we expect constraints to be weaker (stronger) for blue (red) tilted isocurvature primordial power spectra.


\section{Conclusions}
\label{sec:conclusion}

In this work we study the cosmological implications of a DM sector generated through a freeze-in mechanism, a compelling alternative to the traditional WIMP scenario.
In particular we analyze under which conditions a primordial isocurvature perturbation of two classes of freeze-in models would survive the formation of the DM sector.
Because freeze-in DM never achieves thermal equilibrium, an initial isocurvature is never exponentially washed out.

We develop a theoretical framework that allows for the rigorous determination of the background and perturbed macroscopic properties of freeze-in DM, produced through direct decay and 2-to-2 annihilation, starting from the DM PSD.
In particular, we compute for the first time the perturbed PSD from first principles.
We note that this approach does not need to assume how macroscopic properties evolve in time, as in the standard approach where continuity equations are solved to determine the DM properties.
Instead, these properties are fully derived by the evolved PSD and its fluctuations, which are then used to determine the isocurvature perturbation evolution.

Depending on the initial amount of DM generated at the end of inflation and on the amount of primordial isocurvature, we calculate what the final residual isocurvature is at the end of the DM creation epoch.
We forecast the expected sensitivities of next-generation CMB experiments, SO and CMB-S4, in detecting the effects of DM isocurvature on temperature, polarization, and lensing two-point statistics.
We consider different degrees of correlation between the curvature and isocurvature perturbations, showing how future experiments can constraint the presence of an initial freeze-in DM abundance at the percent/subpercent level for a high degree of statistical correlation between curvature and isocurvature perturbations.

Our main findings are that no isocurvature is generated by DM freeze-in. We also find that fluid properties such as shear, pressure, and entropy perturbations are suppressed at large scales, and we quantify the degree of suppression for both the adiabatic and isocurvature modes. In the presence of an initial isocurvature, freeze-in dynamics suppress isocurvature inefficiently, compared to the exponential suppression of thermal DM, which allows an initial isocurvature perturbation to persist. The degree of suppression is well-approximated by background quantities only, with perturbed quantities giving higher order corrections that are suppressed on large scales.  

We present a road map that is applicable to a wide class of DM models. Beyond studying the impact of DM formation on isocurvature, our treatment of analyzing fluctuations from first principles may also be relevant for studying a dark sector that feebly couples to photons, electrons, or neutrinos during the formation of the CMB.
If DM isocurvature is ever detected, the methodology introduced here can serve as an useful tool to connect the dynamics of DM formation with inflationary physics, a rather uncharted territory.


\acknowledgments

We thank Rouven Essig, Daniel Grin, Marc Kamionkowski, and Mauro Valli for comments on the first version of the draft.
KVB thanks Wayne Hu, Gordan Krnjaic, and Albert Stebbins for insightful discussions.
KVB also thanks Pranjal Ralegankar and Daniel Egana-Ugrinovic for comments.
KKB thanks Julian Mu\~{n}oz, Joshua Ruderman, and LianTao Wang for helpful discussions.
The authors further thank Raphael Flauger, Davide Racco, and Antonio Riotto for relevant correspondence.
NB and KKB acknowledge support from the National Science Foundation (NSF) under Grant No.~PHY-2112884.
KVB acknowledges the support of NSF grant PHYS-1915093.


\textbf{Note added:}
In an early version of this work, we found that isocurvature was generated from freeze-in when starting in a flat adiabatic universe. 
This result was an artifact due to a mistake in the numerical implementation, and we correct it here. 
In a previous version of this work, we used $\delta_\mathrm{T} = \widehat{\delta T_0}/\bar{T} \approx \delta_\mathrm{r}/4$, which neglects  changes in $g_{*}(\bar{T})$, instead of $\delta_T = \delta_\mathrm{r}/(4+\frac{d\log g_*}{d\log \bar{T}})$. 
This lead to an only partial cancellation between the second terms in eqs.~\eqref{eq:energy} and \eqref{eq:pertenergy} which artificially contributed a source term for isocurvature on large scales, mimicking DM entropy perturbations caused by entropy changes in the thermal bath. 
Here we correct the previous error, and confirm that freeze-in does not source large isocurvature on scales visible to the CMB $k \, \in \, [10^{-4},10^{-1}]\ \mathrm{Mpc}^{-1}$.


\appendix

\section{Perturbed collision terms}
\label{app:perturbed_collision_terms}

In this appendix we compute the perturbed collision term and its~$\ell$-th moments appearing in the Boltzmann hierarchy.
We also present an analytical approximation of the perturbed energy and momentum exchange terms, appearing in eq.~\eqref{eq:perturbed_fluid_equations1}. 
We perform expansions in terms of Legendre polynomials
\begin{equation}
    P_\ell(\hat{\vec{x}} \cdot \hat{\vec{y}}) = \frac{4\pi}{2\ell+1}\sum^\ell_{m=-\ell} Y_{\ell m}(\hat{\vec{x}}) Y^*_{\ell m}(\hat{\vec{y}}),
\end{equation}
where~$Y_{\ell m}$ are spherical harmonics, which form an orthonormal basis, satisfying
\begin{equation}
    \int d\Omega_{\hat{\vec{n}}} Y_{\ell m}(\hat{\vec{n}}) Y_{\ell' m'}(\hat{\vec{n}}) = \delta^K_{\ell \ell'} \delta^K_{m m'},
\end{equation}
where~$\delta^K$ are Kronecker delta functions.
Due to the orthonormality of the spherical harmonics, we have
\begin{equation}
    \begin{aligned}
        \int \frac{d\Omega_{\hat{\vec{k}}}}{4\pi} P_\ell (\hat{\vec{k}}\cdot\hat{\vec{p}}_1) P_{\ell'} (\hat{\vec{k}}\cdot\hat{\vec{p}}_2) &= \frac{4\pi}{(2\ell+1)(2\ell'+1)}\sum_{m m'} Y^*_{\ell m}(\hat{\vec{p}}_1) Y_{\ell' m'}(\hat{\vec{p}}_2) \int d\Omega_{\hat{\vec{k}}} Y_{\ell m}(\hat{\vec{k}}) Y^*_{\ell' m'}(\hat{\vec{k}}) \\
        &= \frac{\delta^K_{\ell \ell'}}{2\ell+1} P_\ell(\hat{\vec{p}}_1\cdot\hat{\vec{p}}_2).
    \end{aligned}
\end{equation}
We normalize the Legendre polynomials such that~$P_0(x)=1$ and~$P_1(x)=x$.


\subsection{Direct decay}

For the case of freeze-in via decay of a heavy parent particle, the perturbed collision term is
\begin{equation}
    \begin{aligned}
        \widehat{\delta C}(\vec{k}, \vec{p}_\chi, \tau) &= \frac{1}{2} \int \frac{3d^3p_{Z'}}{(2\pi)^3 2E_{Z'}} \int \frac{2d^3p_3}{(2\pi)^32E_3} (2\pi)^4 \delta^{(4)}(p_{Z'}-p_\chi-p_{\overline{\chi}}) \overline{|\mathcal{M}|}^2 \widehat{\delta f}_{Z'}(\tau, \vec{k}, \vec{p}_{Z'}) \\
        &= \frac{g_\chi^2\left(m_{Z'}^2 + 2m_\chi^2\right)}{32 (2\pi)^2} \int \frac{d^3p_{Z'}}{E_{Z'} E_{\overline{\chi}}(|\vec{p}_{Z'}-\vec{p}_\chi|)} \delta^{(1)} \left(E_{Z'}-E_\chi-E_{\overline{\chi}}(|\vec{p}_{Z'}-\vec{p}_\chi|)\right) \widehat{\delta f}_{Z'}(\tau, \vec{k}, \vec{p}_{Z'}) \\
        &= \frac{g_\chi^2\left(m_{Z'}^2 + 2m_\chi^2\right)}{64 \pi p_\chi} \int d\cos\theta_{Z'} dp_{Z'} \frac{p_{Z'}}{E_{Z'}} \delta^{(1)}\left(\cos\theta_{Z'} - \cos\theta^\star\right) \widehat{\delta f}_{Z'}(\tau, \vec{k}, \vec{p}_{Z'}) \, ,
    \end{aligned}
\end{equation}
where~$\cos\theta^\star=(2 E_{Z'} E_\chi - m^2_{Z'}) / (2 p_{Z'} p_\chi)$ and we choose a coordinate system such that
\begin{equation}
    \vec{p}_{Z'} = p_{Z'}(0, \sin\theta_{Z'}, \cos\theta_{Z'}), \quad \vec{p}_\chi = p_\chi(0, 0, 1) \, .
\end{equation}
Under our~$\widehat{\delta T}_\ell (k,p,\tau)\approx \widehat{\delta T}_\ell (k,\tau)$ approximation, the $\ell$-th moment of the collision term is given by
\begin{equation}
    \begin{aligned}
        \widehat{\delta C}_\ell &= i^\ell \int\frac{d\Omega_{\hat{\vec{k}}}}{4\pi} P_\ell (\hat{\vec{k}}\cdot\hat{\vec{p}}_\chi) \widehat{\delta C}(\tau, \vec{k}, \vec{p}_\chi) \\
        &= \frac{g_\chi^2\left(m_{Z'}^2 + 2m_\chi^2\right)}{64 \pi p_\chi} \int d\cos\theta_{Z'} dp_{Z'} \frac{p_{Z'}}{E_{Z'}} \delta^{(1)} \left(\cos\theta_{Z'} - \cos\theta^\star\right) \widehat{\delta f}_\ell(\tau, k, p_{Z'}) P_\ell (\hat{\vec{p}}_{Z'}\cdot\hat{\vec{p}}_\chi) \\
        &= \frac{g_\chi^2\left(m_{Z'}^2 + 2m_\chi^2\right)}{64 \pi p_\chi} \widehat{\delta T}_\ell \int d\cos\theta_{Z'} dp_{Z'} \frac{p_{Z'}}{E_{Z'}} \delta^{(1)} \left(\cos\theta_{Z'} - \cos\theta^\star\right) \frac{E_{Z'}}{\bar{T}^2} \frac{e^{E_{Z'}/\bar{T}}}{(e^{E_{Z'}/\bar{T}}-1)^2} P_\ell (\hat{\vec{p}}_{Z'}\cdot\hat{\vec{p}}_\chi) \, .
    \end{aligned}
\end{equation}
Therefore,
\begin{equation}
    \begin{aligned}
        \widehat{\delta C}_0 &= \frac{g_\chi^2\left(m_{Z'}^2 + 2m_\chi^2\right)\widehat{\delta T}_0}{64 \pi p_\chi} \int dE_{Z'} \frac{E_{Z'}}{\bar{T}^2} \frac{e^{E_{Z'}/\bar{T}}}{(e^{E_{Z'}/\bar{T}}-1)^2} \int d\cos\theta_{Z'} \delta^{(1)}\left(\cos\theta_{Z'} - \cos\theta^\star\right) \\
        &= \frac{g_\chi^2\left(m_{Z'}^2 + 2m_\chi^2\right)\widehat{\delta T}_0}{64 \pi p_\chi} \int^{E_\mathrm{max}}_{E_\mathrm{min}} \frac{dE}{\bar{T}} \frac{E}{\bar{T}} \frac{e^{E/\bar{T}}}{(e^{E/\bar{T}}-1)^2}, \\
        &= \frac{g_\chi^2\left(m_{Z'}^2 + 2m_\chi^2\right)\widehat{\delta T}_0}{64 \pi p_\chi} \left[ \frac{E_\mathrm{min}/\bar{T}}{e^{E_\mathrm{min}/\bar{T}}-1} -\frac{E_\mathrm{max}/\bar{T}}{e^{E_\mathrm{max}/\bar{T}}-1} + \log\frac{1-e^{-E_\mathrm{max}/\bar{T}}}{1-e^{-E_\mathrm{min}/\bar{T}}} \right] \\
    \end{aligned}
\end{equation}
\begin{equation}
    \begin{aligned}
        \widehat{\delta C}_1 &= \frac{g_\chi^2\left(m_{Z'}^2 + 2m_\chi^2\right)\widehat{\delta T}_1}{64 \pi p_\chi} \int dE_{Z'} \frac{E_{Z'}}{\bar{T}^2} \frac{e^{E_{Z'}/\bar{T}}}{(e^{E_{Z'}/\bar{T}}-1)^2} \int d\cos\theta_{Z'} \delta^{(1)} \left(\cos\theta_{Z'} - \cos\theta^\star\right) \cos\theta_{Z'} \\
        &= \frac{g_\chi^2\left(m_{Z'}^2 + 2m_\chi^2\right)\widehat{\delta T}_1}{64 \pi p_\chi} \int^{E_\mathrm{max}}_{E_\mathrm{min}} \frac{dE}{\bar{T}} \frac{E}{\bar{T}} \frac{e^{E/\bar{T}}}{(e^{E/\bar{T}}-1)^2} \frac{2EE_\chi-m^2_{Z'}}{2p_\chi\sqrt{E^2-m^2_{Z'}}}.
    \end{aligned}
\end{equation}

At the level of DM macroscopic properties, the perturbed energy exchange term, corresponding to the~$\ell=0$ multipole, is
\begin{equation}
    \begin{aligned}
        \widehat{\delta \mathcal{Q}}_0 &= \widehat{\delta T}_0 \int \frac{3d^3p_{Z'}}{(2\pi)^3 2E_{Z'}} \frac{2d^3p_\chi}{(2\pi)^3 2E_\chi} \frac{2d^3p_{\overline{\chi}}}{(2\pi)^3 2E_{\overline{\chi}}} E_\chi (2\pi)^4 \delta^{(4)}(p_{Z'}-p_\chi-p_{\overline{\chi}}) \overline{|\mathcal{M}|}^2  \frac{E_{Z'}}{\bar{T}^2} \frac{e^{E_{Z'}/\bar{T}}}{(e^{E_{Z'}/\bar{T}}-1)^2} \\
        &\approx \frac{g^2_\chi (m^2_{Z'}+2m^2_\chi) \widehat{\delta T}_0}{16 (2\pi)^3} \sqrt{1-\frac{4m^2_\chi}{m^2_{Z'}}} \int dE \sqrt{E^2-m^2_{Z'}} \left(\frac{E}{\bar{T}}\right)^{2} \frac{e^{E/\bar{T}}}{(e^{E/\bar{T}}-1)^2} \, , \\
    \end{aligned}
\end{equation}
where we take~$E_\chi \approx E_{Z'}/2$ and use
\begin{equation}
    \int \frac{2d^3p_\chi}{(2\pi)^3 2E_\chi} \frac{2d^3p_{\overline{\chi}}}{(2\pi)^3 2E_{\overline{\chi}}} \delta^{(4)}(p_{Z'}-p_\chi-p_{\overline{\chi}}) = \frac{1}{2\pi} \sqrt{1-\frac{4m_\chi^2}{m^2_{Z'}}} \, .
\end{equation}
For the $\ell=1$ multipole, we have
\begin{equation}
    \begin{aligned}
        \widehat{\delta \mathcal{Q}}_1 &= k\widehat{\delta T}_1 \int \frac{3d^3p_{Z'}}{(2\pi)^3 2E_{Z'}} \frac{2d^3p_\chi}{(2\pi)^3 2E_\chi} \frac{2d^3p_{\overline{\chi}}}{(2\pi)^3 2E_{\overline{\chi}}} p_\chi (2\pi)^4 \delta^{(4)}(p_{Z'}-p_\chi-p_{\overline{\chi}}) \overline{|\mathcal{M}|}^2\\
        &\qquad\qquad\qquad \times\frac{E_{Z'}}{\bar{T}^2} \frac{e^{E_{Z'}/\bar{T}}}{(e^{E_{Z'}/\bar{T}}-1)^2} \frac{2 E_{Z'} E_\chi - m^2_{Z'}}{2 p_{Z'} p_\chi} \\
        &\approx \frac{g^2_\chi (m^2_{Z'}+2m^2_\chi) k \widehat{\delta T}_1}{16 (2\pi)^3} \sqrt{1-\frac{4m^2_\chi}{m^2_{Z'}}} \int dE \left(E^2-m^2_{Z'}\right)^{3/2} \left(\frac{E}{\bar{T}}\right)^2 \frac{e^{E/\bar{T}}}{(e^{E/\bar{T}}-1)^2}, \\
    \end{aligned}
\end{equation}
where we also assume~$p_\chi \approx E_\chi$ and~$p_{Z'}\approx E_{Z'}$.


\subsection{2-to-2 annihilation}

For the case of freeze-in via 2-to-2 annihilation of electron-positron pairs into millicharged DM particles, the perturbed collision term is
\begin{equation}
    \begin{aligned}
        \widehat{\delta C}(\vec{k}, \vec{p}_\chi, \tau) &= \frac{1}{2} \int \frac{2d^3p_{e_-}}{(2\pi)^3 2E_{e_-}} \frac{2d^3p_{e_+}}{(2\pi)^3 2E_{e_+}} \frac{2d^3p_{\overline{\chi}}}{(2\pi)^3 2E_{\overline{\chi}}} (2\pi)^4 \delta^{(4)}(p_{e_-}+p_{e_+}-p_\chi-p_{\overline{\chi}}) \overline{|\mathcal{M}|}^2 \\
        &\qquad\qquad \times \left[\widehat{\delta f}_e(\tau, \vec{k}, \vec{p}_{e_-}) \bar{f}_e(\tau, \vec{p}_{e_+}) + \bar{f}_e(\tau, \vec{p}_{e_-}) \widehat{\delta f}_e(\tau, \vec{k}, \vec{p}_{e_+}) \right] \, .
    \end{aligned}
\end{equation}
The~$\ell$-th moment of the perturbed collision term is thus
\begin{equation}
    \begin{aligned}
        \widehat{\delta C}_\ell &=  \frac{\widehat{\delta T}_\ell}{2} \int \frac{2d^3p_{e_-}}{(2\pi)^3 2E_{e_-}} \frac{2d^3p_{e_+}}{(2\pi)^3 2E_{e_+}} \frac{2d^3p_{\overline{\chi}}}{(2\pi)^3 2E_{\overline{\chi}}} (2\pi)^4 \delta^{(4)}(p_{e_-}+p_{e_+}-p_\chi-p_{\overline{\chi}}) \overline{|\mathcal{M}|}^2 \\
        &\qquad\qquad\qquad\qquad \times \frac{E_{e_-}P_\ell(\hat{\vec{p}}_{e_-}\cdot\hat{\vec{p}}_\chi) + E_{e_+} P_\ell(\hat{\vec{p}}_{e_+}\cdot\hat{\vec{p}}_\chi)}{\bar{T}^2} e^{-(E_{e_-}+E_{e_+})/\bar{T}} \, .
    \end{aligned}
\end{equation}
Therefore, 
\begin{equation}
    \begin{aligned}
        \widehat{\delta C}_0 &= \frac{Q^2_\chi e^4 \widehat{\delta T}_0}{12(2\pi)^3 p_\chi} \int_{s_\mathrm{min}} ds \left[e^{-E_\mathrm{min}/\bar{T}} \left(\frac{E_\mathrm{min}}{\bar{T}}+1\right) - e^{-E_\mathrm{max}/\bar{T}} \left(\frac{E_\mathrm{max}}{\bar{T}}+1\right)\right] \\
        &\qquad\qquad\qquad\qquad\qquad\qquad \times \sqrt{1-\frac{4m_e^2}{s}} \left[1+\frac{2m_e^2}{s}\right] \left[1+\frac{2m_\chi^2}{s}\right], \\
        \widehat{\delta C}_1 &= \frac{Q^2_\chi e^4 \widehat{\delta T}_1}{12(2\pi)^3 p_\chi} \int_{s_\mathrm{min}} ds \left[e^{-E_\mathrm{min}/\bar{T}} \left(\frac{E_\mathrm{min}}{\bar{T}}+1-\frac{s}{2\bar{T}E_\chi}\right) - e^{-E_\mathrm{max}/\bar{T}} \left(\frac{E_\mathrm{max}}{\bar{T}}+1-\frac{s}{2\bar{T}E_\chi}\right)\right]  \\
        &\qquad\qquad\qquad\qquad\qquad\qquad \times \sqrt{1-\frac{4m_e^2}{s}} \left[1+\frac{2m_e^2}{s}\right] \left[1+\frac{2m_\chi^2}{s}\right],
    \end{aligned}
\end{equation}
where for~$\widehat{\delta C}_1$ we assume~$E_{e_-}\approx E_{e_+} \approx p_{e_-} \approx p_{e_+} \approx \bar{T}$ and use~$u = 2m^2_e+2m^2_\chi - t - s$ to approximate
\begin{equation}
    \begin{aligned}
        \frac{E_1\cos\theta_1 + E_2\cos\theta_2}{\bar{T}} &\approx \cos\theta_1+\cos\theta_2 \\
        &= \frac{t - m_1^2 - m_3^2 + 2E_{e_-} E_\chi}{2p_{e_-}p_\chi} + \frac{u - m_2^2 - m_3^2 + 2 E_{e_+} E_\chi}{2p_{e_+}p_\chi} \\
        &\approx \frac{2(E_{e_-}+E_{e_+})E_\chi-s}{2\bar{T}p_\chi}.
    \end{aligned}
\end{equation}

For the perturbed energy and momentum exchange terms, we have
\begin{equation}
    \begin{aligned}
        \widehat{\delta \mathcal{Q}}_0 &= \widehat{\delta T}_0 \int \frac{d^3q_{ee}}{2E_{ee}} ds_{ee} \times I \times \frac{1}{2} \left(\frac{E_{ee}}{\bar{T}}\right)^2 e^{-\frac{E_{ee}}{\bar{T}}} \times J \\
        &\approx \frac{e^4 Q^2_\chi \widehat{\delta T}_0}{6 (2\pi)^5} \int ds_{ee} \sqrt{1-\frac{4m_\chi^2}{s_{ee}}} \sqrt{1-\frac{4m_e^2}{s_{ee}}} \left(1+\frac{2m_e^2}{s_{ee}}\right) \left(1+\frac{2m_\chi^2}{s_{ee}}\right) \\
        &\qquad\qquad\qquad \times \int dE_{ee} \sqrt{E^2_{ee}-s_{ee}} \left(\frac{E_{ee}}{\bar{T}}\right)^2 e^{-\frac{E_{ee}}{\bar{T}}} \\
        &\approx \frac{e^4 Q^2_\chi \widehat{\delta T}_0}{6 (2\pi)^5} \int ds \sqrt{1-\frac{4m_\chi^2}{s}} \sqrt{1-\frac{4m_e^2}{s}} \left(1+\frac{2m_e^2}{s}\right) \left(1+\frac{2m_\chi^2}{s}\right) \\
        &\qquad\qquad\qquad \times \left(6+6\frac{\sqrt{s}}{\bar{T}}+3\frac{s}{\bar{T}^2}+\frac{s^{3/2}}{\bar{T}^3}\right) e^{-\sqrt{s}/\bar{T}}
    \end{aligned}
\end{equation}
and
\begin{equation}
    \begin{aligned}
        \widehat{\delta \mathcal{Q}}_1 &\approx \frac{e^4 Q^2_\chi k \widehat{\delta T}_1}{6 (2\pi)^5} \int ds_{ee} \sqrt{1-\frac{4m_\chi^2}{s_{ee}}} \sqrt{1-\frac{4m_e^2}{s_{ee}}} \left(1+\frac{2m_e^2}{s_{ee}}\right) \left(1+\frac{2m_\chi^2}{s_{ee}}\right) \\
        &\qquad\qquad\qquad \times \int dE_{ee} (E^2_{ee} - s_{ee})^{3/2} e^{-E_{ee}/\bar{T}} \\
        &\approx \frac{e^4 Q^2_\chi k \widehat{\delta T}_1}{6 (2\pi)^5} \int ds \sqrt{1-\frac{4m_\chi^2}{s}} \sqrt{1-\frac{4m_e^2}{s}} \left(1+\frac{2m_e^2}{s}\right) \left(1+\frac{2m_\chi^2}{s}\right) 3 s K_2(\sqrt{s}/\bar{T}) \, ,
    \end{aligned}
\end{equation}
where in both cases we use~$p_\chi \approx E_\chi \approx E_{ee}/2$,
\begin{equation}
    \begin{aligned}
        I &= 4\int \frac{d^3p_\chi}{(2\pi)^3 2E_\chi} \frac{d^3p_{\overline{\chi}}}{(2\pi)^3 2E_{\overline{\chi}}} \delta^{(4)}(q_{ee}-p_\chi-p_{\overline{\chi}}) \approx \frac{1}{(2\pi)^5} \sqrt{1-\frac{4m_\chi^2}{s}}, \\
        J &= \frac{e^4 Q^2_\chi}{6\pi}   \sqrt{1-\frac{4m_e^2}{s}} \left(1+\frac{2m_e^2}{s}\right) \left(1+\frac{2m_\chi^2}{s}\right) \, .
    \end{aligned}
\end{equation}


\section{Fluid perturbations}
\label{app:fluid_perturbations}

By integrating $\ell=0,1,2$ equations of the Boltzmann hierarchy in eqs.~\eqref{eq:boltzmann_hierarchy_synchronous_gauge} over~$2d^3q\epsilon/(2\pi)^3a^4$, $2kd^3q|\vec{q}|/(2\pi)^3a^4$, and $2d^3q|\vec{q}|^2/\left[(2\pi)^3 3\epsilon a^4\right]$, we obtain the evolution equations of the energy density fluctuation~$\delta\rho_\chi$, velocity divergence~$\theta_\chi$, and anisotropic stress~$\sigma_\chi$, respectively:
\begin{equation}
    \begin{aligned}
        &\frac{d\delta\rho_\chi}{d\tau} + 3\mathcal{H}(\delta\rho_\chi+\delta p_\chi) + (\bar{\rho}_\chi+\bar{p}_\chi)\left(\theta_\chi + \frac{1}{2}\frac{dh}{d\tau}\right) = a \delta \mathcal{Q}_{0} \\
        &\frac{d\left[(\bar{\rho}_\chi+\bar{p}_\chi)\theta_\chi\right]}{d\tau} + 4\mathcal{H}(\bar{\rho}_\chi+\bar{p}_\chi)\theta_\chi - k^2\delta p_\chi + k^2(\bar{\rho}_\chi+\bar{p}_\chi)\sigma_\chi = a\delta \mathcal{Q}_{1} \\
        &\frac{d\left[(\bar{\rho}_\chi+\bar{p}_\chi)\sigma_\chi\right]}{d\tau} + \mathcal{H}(\bar{\rho}_\chi+\bar{p}_\chi)\left(5\sigma_\chi-\tilde{\Sigma}_\chi\right) - \frac{4}{15}(\bar{\rho}_\chi+\bar{p}_\chi)\tilde{\Theta}_\chi \\
        &\qquad\qquad\qquad\qquad\qquad\qquad - \left(\frac{2}{15}\frac{dh}{d\tau}+\frac{4}{5}\frac{d\eta}{d\tau}\right)(5\bar{p}_\chi - \tilde{p}_\chi) = 0 \, ,
    \end{aligned}
\label{eq:perturbed_fluid_equations}
\end{equation}    
where we introduce a higher moment of pressure~$\tilde{p}_\chi$, velocity divergence~$\tilde{\Theta}_\chi$, and anisotropic stress~$\tilde{\Sigma}_\chi$ defined as~\cite{ma:perturbationtheory, shoji:highermoments, lesgourgues:classncdm}
\begin{align}
    \tilde{p}_\chi &= \frac{1}{a^4} \int \frac{2d^3q}{(2\pi)^3}\frac{q^4}{3\epsilon^3} \bar{f}_\chi \\
    (\bar{\rho}_\chi+\bar{p}_\chi)\tilde{\Theta}_\chi &= \frac{k}{a^4} \int \frac{2d^3q}{(2\pi)^3} \frac{q^3}{\epsilon^2}  \widehat{\delta f}_{\chi,1} \\
    \quad (\bar{\rho}_\chi+\bar{p}_\chi)\tilde{\Sigma}_\chi &= \frac{1}{a^4} \int \frac{2d^3q}{(2\pi)^3} \frac{2q^4}{3\epsilon^3} \widehat{\delta f}_{\chi,2}\, .
\end{align}
For relativistic DM with~$q^2\sim \epsilon^2$, we can safely make the approximations~$\tilde{p}_\chi\sim \bar{p}_\chi$, $\tilde{\Theta}_\chi\sim \theta_\chi$, and~$\tilde{\Sigma}_\chi\sim \sigma_\chi$, thus recovering eq.~\eqref{eq:perturbed_fluid_equations1}.


\section{Kodama-Sasaki notation}
\label{app:kodama_sasaki_notation}

It is useful to map equations~\eqref{eq:perturbed_fluid_equations} into the same equations in the Kodama-Sasaki (KS) notation~\cite{kodama:gaugeinvariantperturbations}.
In KS notation, the evolution equations for any given species, without specifying any gauge, are
\begin{equation}
    \begin{aligned}
        \delta \rho_\alpha ' &+ 3\mathcal{H} \left( \delta\rho_\alpha + \bar{p}_\alpha \pi_{L\alpha} \right) + (\bar{\rho}_\alpha + \bar{p}_\alpha ) ( k v_\alpha + 3 H'_L ) = a (\varepsilon_\alpha - A) \bar{\mathcal{Q}}_\alpha \\
        \left[\left(\bar{\rho}_\alpha+\bar{p}_\alpha\right) \left(v_\alpha-B\right)\right]' &+ 4\mathcal{H}\left(\bar{\rho}_\alpha+\bar{p}_\alpha\right) \left(v_\alpha-B\right) - k\bar{p}_\alpha \pi_{L\alpha} - k\left(\bar{\rho}_\alpha+\bar{p}_\alpha\right)A \\
        \qquad\qquad\qquad &+ \frac{2}{3}k\bar{p}_\alpha\pi_{T\alpha} = a\bar{\mathcal{Q}}_\alpha \left(v-B\right) + a\mathcal{H}\left(\bar{\rho}_\alpha+\bar{p}_\alpha\right)f_\alpha \, .
    \end{aligned}
\end{equation}
Matching notations, we have~$v_\alpha = \theta_\alpha/k$, $\pi_{L\alpha} = \delta p_\alpha / \bar{p}_\alpha$ and~$\pi_{T\alpha}=\frac{3}{2}\left(\bar{\rho}_\alpha+\bar{p}_\alpha\right)\sigma_\alpha/\bar{p}_\alpha$. 
Therefore, in the case of DM in the synchronous comoving gauge~$(A=B=0,H_L=h/6)$, we have
\begin{equation}
    \begin{aligned}
        \delta \rho_\chi ' &+ 3\mathcal{H} \left( \delta\rho_\chi + \delta p_\chi \right) + (\bar{\rho}_\chi + \bar{p}_\chi ) \left( \theta_\chi + \frac{h'}{2} \right) = a \varepsilon_\chi \bar{\mathcal{Q}}_\chi, \\
        \left[\left(\bar{\rho}_\chi+\bar{p}_\chi\right) \theta_\chi\right]' &+ 4\mathcal{H}\left(\bar{\rho}_\chi+\bar{p}_\chi\right) \theta_\chi - k^2\delta p_\chi + k^2\left(\bar{\rho}_\chi+\bar{p}_\chi\right)\sigma_\chi = a\bar{\mathcal{Q}}_\chi \theta + a\mathcal{H}k\left(\bar{\rho}_\chi+\bar{p}_\chi\right)f_\chi,
    \end{aligned}
\end{equation}
from which we derive that
\begin{equation}
    \varepsilon_\chi = \frac{\delta \mathcal{Q}_{0}}{\bar{\mathcal{Q}}_\chi}, \qquad f_\chi = \frac{\delta \mathcal{Q}_{1}-\bar{\mathcal{Q}}_\chi\theta}{\mathcal{H}k(\bar{\rho}_\chi+\bar{p}_\chi)}.
\end{equation}
The conservation of the stress-energy tensor imposes the constraints~$\bar{\mathcal{Q}}_\mathrm{r}=-\bar{\mathcal{Q}}_\chi$, $\varepsilon_\mathrm{r}=\varepsilon_\chi$, and~$f_\mathrm{r}=-(\bar{\rho}_\chi+\bar{p}_\chi)f_\chi/(\bar{\rho}_\mathrm{r}+\bar{p}_\mathrm{r})$.
Finally, under a gauge transformation characterized by the two free functions~$T$ and $L$~\cite{kodama:gaugeinvariantperturbations}, the metric scalar variables transform as
\begin{equation}
    \tilde{A}=A-T'-\mathcal{H}T,\quad \tilde{B}=B+L'+kT,\quad \tilde{H}_L=H_L-(k/3)L-\mathcal{H}T,\quad \tilde{H}_T=H_T+kL,
\end{equation}
while the species fluctuations and energy-momentum four-vector transform as
\begin{equation}
\label{eq:transforms}
    \begin{aligned}
        \tilde{\delta}_\alpha &= \delta_\alpha+3\mathcal{H}(1+w_\alpha)(1-q_\alpha)T,\quad \tilde{v}_\alpha = v_\alpha+L', \\
        \tilde{\pi}_L &= \pi_L + 3 \frac{c^2_{a,\alpha}}{w_\alpha} (1+w_\alpha) (1-q_\alpha) \mathcal{H} T,\quad \tilde{\pi}_T = \pi_T, \\
        \tilde{\varepsilon}_\alpha &= \varepsilon_\alpha - \frac{\bar{\mathcal{Q}}'_\alpha}{\mathcal{H}\bar{\mathcal{Q}}_\alpha} \mathcal{H}T, \quad \tilde{f}_\alpha = f_\alpha,
    \end{aligned}
\end{equation}
respectively.  
Given these definitions, we verify the gauge invariance of the isocurvature perturbation we employ in this work.


\section{Comment on the KS formalism}
\label{app:comment}

Alternative definitions of the isocurvature perturbation have been proposed in the literature (e.g., see refs.~\cite{kodama:gaugeinvariantperturbations, Malik:2001rm, malik:gaugeinvariantperturbations, malik:gaugeinvariantperturbeationsreview}). 
In this appendix, we compare definitions and comment on their interpretations.

In the KS formalism, the amplitude of the isocurvature mode [cf., eq.~\eqref{eq:isocurvature}] has a simple interpretation in the comoving orthogonal gauge, in terms of the background energy or number density and its fluctuation:
\begin{equation}
    S^\mathrm{KS}_{\chi \mathrm{r}} = \frac{\delta \rho_{\chi}}{\bar{\rho}_{\chi}(1+w_{\chi})} -\frac{\delta \rho_{\text{r}}}{\bar{\rho}_\mathrm{r}(1+w_{\text{r}})  } = \frac{\delta n_{\chi}}{\bar{n}_{\chi}} - \frac{\delta n_{\text{r}}}{\bar{n}_{\text{r}}} \, .
\label{eq:SKS}
\end{equation}
for standard non-interacting DM and radiation sectors.
In the alternative formalism introduced by Malik and Wands (MW)~\cite{Malik:2001rm, malik:gaugeinvariantperturbations, malik:gaugeinvariantperturbeationsreview}, the amplitude is instead defined to be
\begin{equation}
    S^{\text{MW}}_{\chi \mathrm{r}} = - 3\mathcal{H} \left(  \frac{\delta \rho_{\chi}}{d\bar{\rho}_{\chi}/d\tau} -  \frac{\delta \rho_{\text{r}}}{d\bar{\rho}_\mathrm{r}/d\tau} \right)  = \frac{\delta \rho_{\chi}}{\bar{\rho}_{\chi} (1-\mathtt{q}_{\chi}) (1+w_{\chi})} -\frac{\delta\rho_\mathrm{r}}{\bar{\rho}_\mathrm{r} (1-\mathtt{q}_\mathrm{r}) (1+w_\mathrm{r})} \,  .
\label{eq:SMW}
\end{equation}
This latter definition ensures that if the single clock condition in eq.~\eqref{eq:single_clock} is fulfilled, then~$S^\mathrm{MW}_{\chi \mathrm{r}}=0$.
In the absence of energy transfer, the definitions from KS and MW coincide~\cite{Malik:2001rm, malik:gaugeinvariantperturbations, malik:gaugeinvariantperturbeationsreview}.
However, when~$\mathtt{q}_\chi,\mathtt{q}_\mathrm{r}\neq 0$, the two definitions differ by 
\begin{equation}
   S^{\text{MW}}_{\chi \mathrm{r}} - S^\mathrm{KS}_{\chi \mathrm{r}} \approx \frac{3}{4}\frac{\mathtt{q}_{\chi}}{1- \mathtt{q}_{\chi}} \Delta\, \propto x^2 \ll 1
\end{equation}
during radiation domination ($\mathtt{q}_\mathrm{r}\ll 1,\ \bar{\rho}_\chi / \bar{\rho}_\mathrm{r}\ll 1$).

Both definitions of isocurvature in eq.~\eqref{eq:isocurvature} and eq.~\eqref{eq:SMW} are gauge invariant\footnote{It has been claimed that the isocurvature definition of Kodama and Sasaki is not gauge invariant~\cite{Malik:2001rm, Racco1}. 
To clarify, eq.~\eqref{eq:isocurvature}, the definition used by Kodama and Sasaki, is gauge invariant, as can be seen by using the transformations in eq.~\eqref{eq:transforms}; however, the definition in terms of perturbed number densities in eq.~\eqref{eq:SKS} is not.} and thus take the same value in every coordinate system \cite{Hu:2003hjx}.
However, the KS definition $S^\mathrm{KS}_{\chi \mathrm{r}}$ is constructed as a linear combination of gauge-dependent perturbations~\cite{1992ApJ...395...34B}, which obfuscates its interpretation in terms of physical quantities; the interpretation only becomes apparent in gauges with a particular choice of physical time slicing, such as comoving orthogonal gauge. 
On the other hand, the MW definition of isocurvature has a clear geometric interpretation \cite{1992ApJ...395...34B}.
In the large scale limit $k \to 0$, which is relevant for CMB observations, the evolution of $S^\mathrm{KS}_{\chi \mathrm{r}}$ coincides with $S^{\text{MW}}_{\chi \mathrm{r}}$; however, there are differences at finite $k$. 
The redefinition of the isocurvature variable by MW changes the evolution eq.~\eqref{eq:KSfull3} for the amplitude of the isocurvature $S_{\chi \mathrm{r}}^{\text{MW}}$, such that the source terms proportional to the adiabatic perturbation $\Delta$ vanish, both in eq.~\eqref{eq:KSfull3} explicitly and in the energy exchange term $E_{\chi \mathrm{r}}$ in eq.~\eqref{eq:Eexchange}~\cite{Malik:2001rm}. 
The remaining energy exchange term sourcing isocurvature, $E^{\text{MW}}_{\chi \mathrm{r}}$, can be written explicitly in terms of the non-adiabatic part of the perturbed energy transfer:
\begin{equation}
    E^{\text{MW}}_{\chi \mathrm{r}} \propto \left(\delta \mathcal{Q} -\frac{\mathcal{Q}'}{\rho_{\chi}'} \delta \rho_{\chi}\right) \, .
\end{equation}
In the MW formalism, it is then straightforward to show that the non-adiabatic part of the perturbed energy transfer is zero in a universe with zero initial isocurvature, $S^{\text{MW}}_{\chi \mathrm{r}} = 0$, given that the energy transfer is a function of the radiation density $\mathcal{Q}(\rho_r)$, such that $\delta \mathcal{Q} = (d\mathcal{Q}/d\tau)/(d\rho_\mathrm{r}/d\tau) \times \delta \rho_\mathrm{r}$ during freeze-in. 
This argument has been presented in \cite{Strumia:2022qvj, Racco1}.
Yet, in the KS formalism, source terms proportional to the adiabatic perturbation $\Delta$ remain, yielding nonzero, scale-dependent contributions to isocurvature on super-horizon scales. 
This effect is not observable: it is an artifact of the KS formalism introducing a small mixing between the curvature and isocurvature perturbations.
Regardless, both formalisms are consistent with each other, since differences are highly suppressed.

In conclusion, on large scales both formalisms are consistent with each other since differences are highly suppressed, thus our results from section~\ref{subsec:initial_isocurvature_evolution} are applicable to describe the amplitude of the isocurvature mode constrained by the \textit{Planck} satellite.


\bibliography{biblio}
\bibliographystyle{utcaps}

\end{document}